%% file: arxiv.tex
\documentclass[10pt,conference,screen]{IEEEtran}
\usepackage[switch]{lineno}
\usepackage{cuted}
\newcommand{\toolname}{\textsc{EvoRepair}\xspace}
\usepackage{mdframed}
\usepackage{bbding}

\usepackage{tcolorbox}
\tcbuselibrary{breakable}
\usepackage{pifont}
\usepackage{enumerate}
\usepackage{ragged2e}
\usepackage{makecell}
\usepackage{stfloats}
\usepackage{amsmath}
\usepackage[vlined, ruled]{algorithm2e}
\usepackage{algorithmic}
\usepackage{textcomp}
\usepackage{multirow}
\usepackage{subfigure}
\usepackage{diagbox}
\usepackage{listings}
\usepackage{lstlinebgrd}
\usepackage{float}
\usepackage{url}
\usepackage{xspace}
\usepackage{booktabs}
\usepackage{multirow}
\usepackage{makecell}
\usepackage{siunitx}
\usepackage{xcolor}
\usepackage[hidelinks]{hyperref}
\usepackage{wrapfig}
\IEEEoverridecommandlockouts
\usepackage{orcidlink}
\usepackage{enumitem}
\usepackage[T1]{fontenc}
\definecolor{deepblue}{rgb}{0,0,0.5}
\definecolor{deepgreen}{rgb}{0,0.5,0}
\definecolor{deepred}{rgb}{0.6,0,0}
\definecolor{darkorange}{RGB}{255,140,0}
\definecolor{lightgray}{rgb}{0.93,0.93,0.93}
\definecolor{deepgray}{rgb}{0.25,0.25,0.25}

\tcbuselibrary{minted,skins}
\lstdefinelanguage{Java}{
	basicstyle=\small\ttfamily,
	numberstyle=\color{deepgray},
	stepnumber=1,
	numbersep=8pt,
	showstringspaces=false,
	breaklines=true,
	frame=lines,
	backgroundcolor=\color{lightgray},
	commentstyle=\color{deepgreen},
	keywordstyle=\color{deepblue},
	stringstyle=\color{deepred},
	tabsize=4,
	captionpos=b,
	morekeywords={public, class, void, int, if, else, for, while, return, true, false},
	emph={String, System},
	emphstyle=\color{darkorange},
	alsoletter={.,;:[]()},
}
\definecolor{codebg}{RGB}{248,249,251}
\definecolor{codeframe}{RGB}{185,194,205}
\definecolor{codekw}{RGB}{42,89,103}
\definecolor{codecomment}{RGB}{63,102,82}
\definecolor{codestring}{RGB}{122,75,43}
\definecolor{vulnhl}{RGB}{255,235,235}
\definecolor{exphl}{RGB}{234,242,247}
\definecolor{patchhl}{RGB}{228,246,236}
\lstdefinelanguage{JavaScript}{
  keywords={const,var,function,return,if,require},
  keywordstyle=\bfseries\color{codekw},
  sensitive=true,
  morecomment=[l]{//},
  morecomment=[s]{/*}{*/},
  commentstyle=\itshape\color{codecomment},
  morestring=[b]",
  morestring=[b]',
  stringstyle=\color{codestring}
}
\lstdefinestyle{motivationcode}{
  language=JavaScript,
  basicstyle=\footnotesize\ttfamily,
  numbers=none,
  frame=single,
  framerule=0.35pt,
  rulecolor=\color{codeframe},
  backgroundcolor=\color{codebg},
  showstringspaces=false,
  breaklines=true,
  breakatwhitespace=true,
  columns=fullflexible,
  keepspaces=true,
  tabsize=2,
  xleftmargin=0.4em,
  framexleftmargin=0.4em,
  linebackgroundcolor={%
    \ifnum\value{lstnumber}=3\color{vulnhl}\fi
    \ifnum\value{lstnumber}=8\color{vulnhl}\fi
    \ifnum\value{lstnumber}=14\color{exphl}\fi
    \ifnum\value{lstnumber}=18\color{patchhl}\fi
    \ifnum\value{lstnumber}=19\color{patchhl}\fi
    \ifnum\value{lstnumber}=20\color{patchhl}\fi
    \ifnum\value{lstnumber}=21\color{patchhl}\fi
    \ifnum\value{lstnumber}=26\color{patchhl}\fi
  },
  linebackgroundsep=0pt,
  aboveskip=2pt,
  belowskip=2pt
}
\usepackage[utf8]{inputenc}
\usepackage{graphicx}
\usepackage{color,xcolor}
\usepackage{url}
\usepackage{stmaryrd}
\usepackage{verbatimbox}
\usepackage{bm}
\usepackage[justification=centering]{caption}

\usepackage{makecell}
\usepackage{booktabs}

\usepackage{colortbl}

\usepackage[normalem]{ulem}

\newcommand{\delete}[1]{\iffalse{#1}\fi}

\AtBeginDocument{%
  \providecommand\BibTeX{{%
    \normalfont B\kern-0.5em{\scshape i\kern-0.25em b}\kern-0.8em\TeX}}}
    
\begin{document}

\title{Learning from Repair Experience: Self-Evolving Agents for Automated Vulnerability Repair}

\author{
\IEEEauthorblockN{
Haichuan Hu\textsuperscript{1}\orcidlink{0009-0002-3007-488X},
Guoqing Xie\textsuperscript{2}\orcidlink{0009-0002-3475-0021},
Quanjun Zhang\textsuperscript{1,*}\orcidlink{0000-0002-2495-3805},
Jiawei Liu\textsuperscript{2}\orcidlink{0000-0002-4930-9637},\\
Shengcheng Yu\textsuperscript{3}\orcidlink{0000-0003-4640-8637},
Chunrong Fang\textsuperscript{2}\orcidlink{0000-0002-9930-7111},
Zhenyu Chen\textsuperscript{2}\orcidlink{0000-0002-9592-7022},
Liang Xiao\textsuperscript{1}\orcidlink{0000-0003-0178-9384}
}
\IEEEauthorblockA{
\textsuperscript{1}Nanjing University of Science and Technology,
Nanjing, China\\
\textsuperscript{2}Nanjing University,
Nanjing, China\\
\textsuperscript{3}Technical University of Munich,
Munich, Germany\\[3pt]
\textit{huhaichuan2024@gmail.com,
522025320181@smail.nju.edu.cn}\\
\textit{quanjunzhang@njust.edu.cn,
jwliu@nju.edu.cn}\\
\textit{shengcheng.yu@tum.de,
fangchunrong@nju.edu.cn}\\
\textit{zychen@nju.edu.cn,
xiaoliang@mail.njust.edu.cn}
}
\thanks{\textsuperscript{*}Corresponding author:
Quanjun Zhang (quanjunzhang@njust.edu.cn).}
}

\IEEEaftertitletext{\vspace{-24pt}}
\maketitle

\begin{abstract}

Large Language Models (LLMs) have shown promise for automated vulnerability repair (AVR), but they still face several limitations, including the lack of intra-vulnerability experience accumulation and the lack of cross-vulnerability experience reuse. As a result, LLMs may repeatedly make similar mistakes during iterative repair and underutilize valuable repair knowledge from historical vulnerabilities.
To address these challenges, we propose \toolname{}, an experience-based self-evolving AVR agent framework that enables LLMs to accumulate, refine, and leverage domain-specific knowledge across long-horizon vulnerability repairs. \toolname{} follows a cyclic learn-and-repair process that retrieves relevant past experiences to guide repair, extracts new experiences from repair trajectories, and updates an experience bank using quality-aware scoring. We evaluate \toolname{} against 15 representative vulnerability repair baselines on PATCHEVAL and SEC-bench. Results show that \toolname{} achieves the best overall performance, reaching 93.47\% on PATCHEVAL, 87.00\% on SEC-bench, and 90.46\% overall. In particular, \toolname{} outperforms the latest LLM-based baseline LoopRepair by 39.56\% and 33.50\% on PATCHEVAL and SEC-bench, respectively, and surpasses IntentFix by 70.86\% and 50.50\%. Across both benchmarks, \toolname{} also exceeds the recent self-evolving agent Live-SWE-Agent by 6.98\% overall. Additional transfer experiments on VUL4J further demonstrate the robustness of \toolname{} across models, programming languages, and datasets. These findings demonstrate that experience-based self-evolution substantially strengthens agentic AVR and goes beyond existing self-evolving techniques. 

\end{abstract}


\section{Introduction}
\begin{figure}[t]
\centering
\includegraphics[width=0.95\linewidth]{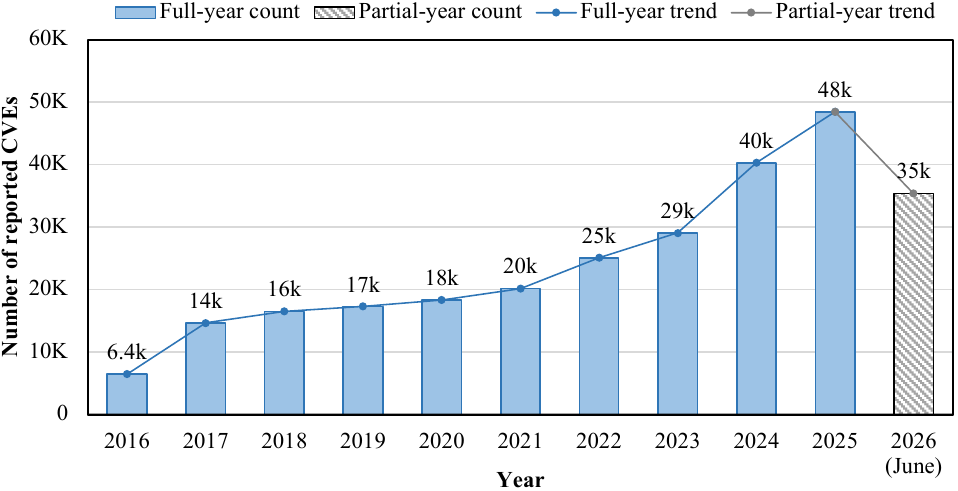}
\caption{Yearly growth in reported CVEs.}
\label{fig:cve_count_by_year}
\end{figure}

As shown in Figure~\ref{fig:cve_count_by_year}, the number of reported CVEs accelerated after 2019; in 2025 alone, more than 48,000 new CVEs were reported, representing a 20\% year-over-year increase.

Given the growing number of disclosed vulnerabilities, Automated Vulnerability Repair (AVR)~\cite{zhang2023pre,shen2020survey,hu2026psearch,hu2026multifixer} has emerged as a promising approach for accelerating vulnerability fix and reducing security risks. Early work on AVR~\cite{gao2019crash,gao2021beyond}, predominantly based on program analysis and search-based methods, has shown effectiveness in synthesizing patches for certain types of vulnerabilities, albeit typically within constrained domains. 
Subsequent learning-based AVR techniques~\cite{fu2023toward,zhang2024vuladvisor,zhang2023pre} fine-tune pre-trained models on large-scale vulnerability datasets to capture diverse repair patterns. 
However, learning-based methods face an important evaluation challenge.
Prior work generally adopts CodeBLEU~\cite{ren2020codebleumethodautomaticevaluation} and Exact Match as evaluation metrics, such execution-agnostic measures are insufficient for assessing real-world repair effectiveness. 
Specifically, they primarily capture surface-level similarity to reference patches and may therefore overestimate a model's ability to generalize across heterogeneous datasets.

Recently, the rise of Large Language Models (LLMs) has opened new opportunities for AVR~\cite{zhou2025large}.
Compared with earlier learning-based methods, LLM-based approaches~\cite{costin2024evaluating,bao2025smart,shao2026fix} can achieve strong repair performance under zero-shot or few-shot settings. Moreover, they make end-to-end repair feasible and use test feedback to iteratively revise incorrect patches, reducing reliance on the matching-based evaluation schemes commonly used in prior learning-based methods. Although LLM-based methods represent the state of the art in AVR, they still face two technical challenges.
(1) \textbf{Lack of intra-vulnerability repair experience accumulation.} Existing AVR methods make limited use of the experience generated during the repair of a single vulnerability. Although multiple repair trajectories may be explored, intermediate successes, failures, and diagnostic signals are rarely distilled into reusable experience for subsequent attempts. Here, we define \emph{experience} as structured repair knowledge abstracted from repair trajectories, rather than raw interaction logs or one-off patches. Without such experience accumulation, LLMs may repeatedly follow unproductive repair paths, make similar mistakes across trajectories, and miss opportunities to iteratively refine its repair strategy. This stateless repair behavior reduces efficiency and weakens LLMs' ability to improve within the current task.
(2) \textbf{Lack of cross-vulnerability repair experience reuse.}
Existing AVR methods~\cite{ye2025well,fakih2025llm4cve} largely treat each vulnerability as an isolated repair task and do not systematically reuse repair experience across different vulnerabilities.
This limitation becomes increasingly critical as the number of reported vulnerabilities continues to rise, as shown in Figure~\ref{fig:cve_count_by_year}, making one-by-one isolated repair progressively less scalable.
Because CWE and CVE provide structured taxonomies, vulnerabilities within the same or related categories often share similar causes, contexts, and repair patterns.
Although retrieval-augmented approaches~\cite{cheng2025automated} that retrieve reference patches or examples can provide useful guidance, they remain limited in AVR because the retrieved knowledge is typically static and instance-specific. In contrast, effective AVR requires repair experience that is reusable, generalizable, and continuously refined through a self-evolving repair process.
Consequently, relying solely on static retrieval limits cross-case generalization and hinders the accumulation of transferable repair knowledge for large-scale vulnerability repair.

Together these challenges indicate that effective AVR requires not only generating patches for the current vulnerability, but also accumulating, refining, and reusing repair experience both within and across vulnerabilities. This motivates the following research question: \textbf{\textit{how can AVR systems effectively accumulate and reuse repair experience both across vulnerabilities and within a single vulnerability?}}
To answer this question, we propose \toolname{}, an experience-based self-evolving framework for AVR. The core idea of \toolname{} is a cyclic two-stage learn-and-repair paradigm. In the learning stage, \toolname{} extracts domain-specific repair knowledge from historical vulnerability repair trajectories and stores it as long-term repair experience. In the repair stage, \toolname{} retrieves relevant experiences for similar vulnerabilities and uses them to guide the repair process toward more promising repair directions. By alternating between these two stages, \toolname{} continuously improves both repair effectiveness and the quality of the accumulated repair experience.

We evaluate \toolname{} against 15 representative AVR baselines on PATCHEVAL~\cite{wei2025patcheval} and SEC-bench~\cite{lee2025sec}. Overall, the results show that \toolname{} achieves the best performance across both datasets, reaching 93.47\% on PATCHEVAL, 87.00\% on SEC-bench, and 90.46\% overall. In particular, \toolname{} outperforms the strongest LLM-based baseline, LoopRepair, by 39.56\% and 33.50\% on PATCHEVAL and SEC-bench, respectively, and surpasses IntentFix by 70.86\% and 50.50\%. \toolname{} also exceeds the recent self-evolving SE agent Live-SWE-Agent by 6.98\% overall. Furthermore, cross-dataset transfer experiments on VUL4J demonstrate that the experiences synthesized by \toolname{} generalize across different datasets, programming languages, and model backbones.

In summary, we make the following contributions:

\begin{itemize}[leftmargin=*]
    \item \textbf{Experience-Driven Self-Evolving AVR.}
We formulate AVR as an experience-driven test-time learning problem, where repair trajectories are treated as reusable supervision for continuously improving vulnerability-repair agents.

\item \textbf{Novel Method.}
We propose \toolname{}, a cyclic learn-and-repair framework that operationalizes this formulation through structured experience extraction, quality-aware scoring, experience-bank updating, and retrieval-guided repair.
    
    \item \textbf{Extensive Experiments.} We evaluate \toolname{} against 15 representative vulnerability repair baselines on PATCHEVAL and SEC-bench, and assess the transfer ability of \toolname{} on VUL4J. The results demonstrate the strong effectiveness of \toolname{}, as well as its cross-dataset transferability, cross-language generality, and applicability across different backbone models.
\end{itemize}

\section{Background and Motivation}
\subsection{Automated Vulnerability Repair}

Automated Vulnerability Repair~\cite{DBLP:conf/icse/NollerS0R22,DBLP:journals/tdsc/BellevilleSVAF21} aims to fix security flaws with minimal human intervention. A typical AVR pipeline includes vulnerability detection, patch generation, verification, and deployment. Vulnerable locations can be identified using static-analysis tools such as Infer~\cite{infer} and SpotBugs~\cite{spotbugs}, or model-based methods. Candidate patches are then synthesized~\cite{DBLP:journals/tse/ChenKM23} and validated through recompilation, functional testing, and security re-analysis before deployment~\cite{DBLP:journals/corr/abs-2203-05166}.
Existing AVR methods can be broadly grouped into three categories. Early NMT-based approaches formulate vulnerability repair as translation from vulnerable code to repaired code; representative examples include VRepair~\cite{DBLP:journals/tse/ChenKM23} and SeqTrans~\cite{DBLP:journals/tse/ChiQLZY23}. Pretraining-based methods further improve repair performance by fine-tuning pretrained code models on vulnerability datasets. For example, VulRepair~\cite{fu2022vulrepair} fine-tunes CodeT5 on CVEFixes~\cite{DBLP:journals/corr/abs-2107-08760}, while VulMaster~\cite{10.1145/3597503.3639222} extends CodeT5 with abstract syntax trees and CWE examples. More recently, LLM-based approaches have become dominant because they support end-to-end vulnerability localization, patch generation, and validation without task-specific training~\cite{DBLP:conf/kbse/WenLYGY25,DBLP:journals/tosem/ZhouCSL25,hu2025can,hu2025repair}. Representative examples include LLM4CVE~\cite{DBLP:conf/dsd/FakihDBAOSF25} and VulnRepairEval~\cite{DBLP:journals/corr/abs-2509-03331}. Among them, agent-based methods currently represent the strongest line of work by augmenting LLMs with memory and external tools~\cite{DBLP:journals/corr/abs-2504-07634,DBLP:journals/corr/abs-2601-13933,kim2026patchislandorchestrationllmagents}. For example, VulnResolver~\cite{DBLP:journals/corr/abs-2601-13933} adopts a hybrid multi-agent design, while PatchIsland~\cite{kim2026patchislandorchestrationllmagents} integrates agent orchestration with continuous fuzzing pipelines. However, existing agent-based AVR systems still rely heavily on the parametric knowledge of the underlying LLM and do not explicitly accumulate and reuse repair experience across tasks. This limitation motivates our study of self-evolving vulnerability-repair agents.

\subsection{Self-Evolving Agents}
Self-evolving techniques~\cite{zhou2026self} aim to improve agents autonomously through interaction experience with minimal manual supervision. Expel~\cite{zhao2024expel} introduces experiential learning by extracting natural-language knowledge from prior interactions to support future decision-making. EvolveR~\cite{wu2025evolver} extends this idea by formalizing a full experience lifecycle that distills multi-turn interactions into reusable principles and closes the self-improvement loop through reinforcement learning. Absolute Zero Reasoner~\cite{zhao2025absolute}, R-Zero~\cite{huang2025r}, and Multi-Agent Evolve~\cite{chen2025multi} further advance this line of work through verifiable feedback, co-evolutionary learning, and structured multi-agent pipelines.
Recently, self-evolving mechanisms have been applied to code agents in software engineering~\cite{lin2025se,xia2025live,hu2026controlled,ding2026swe,weng2026group,chen2025swe,hao2026recreate,zhang2026effectiveness,zhang2024machine,zhang2026compass}, yielding strong progress in program repair. SWE-EXP~\cite{chen2025swe} extracts experiences offline for later retrieval, while Live-SWE-Agent~\cite{xia2025live} evolves tools during repair. AgentMem~\cite{xu2025amem} updates memory through semantic association and retrieves by relevance. These approaches provide foundations for knowledge reuse, but vulnerability repair requires security-specific reasoning beyond the issue repair tasks commonly evaluated on SWE-bench~\cite{jimenez2023swe}. Building on these foundations, \toolname{} contributes an AVR-specific experience lifecycle that compares experience quality across repair turns and retrieves guidance using both relevance and quality. Its five-dimensional schema captures vulnerability analysis, repair rationale, trajectory analysis, reusable rules, and reflection, while CVE/CWE matching and disclosure-time constraints support relevant, chronologically valid reuse. This integration enables experience refinement within a vulnerability and knowledge reuse across vulnerabilities.

\subsection{Motivation}

In complex problem-solving tasks, success often depends on whether a system can leverage prior experience rather than rely solely on stochastic trial-and-error. Without mechanisms to summarize past failures and reuse successful strategies, repeated attempts tend to yield diminishing returns.
We observe the same pattern in AVR. When an LLM-based agent fails to identify a promising repair path early, simply increasing the number of repair steps or interaction turns usually leads to much higher cost with only limited gains. This blind scaling often causes the agent to repeat similar reasoning and repair mistakes. To address this limitation, we shift the repair process from exhaustive trial-and-error to experience-guided refinement. By maintaining a dynamic experience bank that records successful repair patterns and common failure modes, the agent can retrieve relevant prior knowledge, avoid redundant exploration, and improve both repair efficiency and repair quality.

\begin{figure}[t]
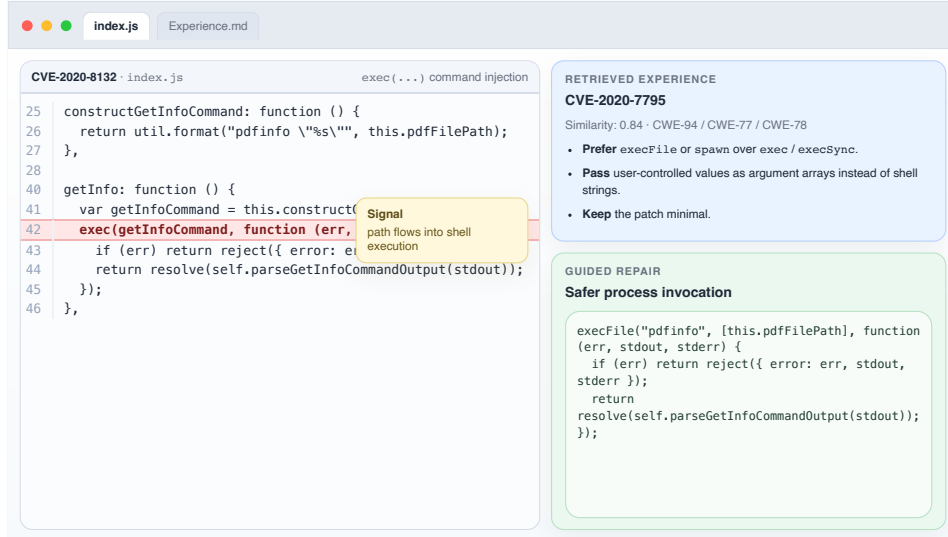

\begin{lstlisting}[style=motivationcode]
/* [VULNERABLE VERSION] path -> shell command. */
constructGetInfoCommand: function () {
  return 'pdfinfo ' + this.pdfFilePath;
}

getInfo: function () {
  var cmd = this.constructGetInfoCommand();
  exec(cmd, function (err, stdout) {
    if (err) return reject({ error: err });
    return resolve(parse(stdout));
  });
}

/* [EXPERIENCE] Use argv arrays, not shell strings... */

/* [CORRECT PATCH] command and arguments are separated. */
constructGetInfoCommand: function () {
  return {
    cmd: 'pdfinfo',
    args: [this.pdfFilePath]
  };
}

getInfo: function () {
  var cmd = this.constructGetInfoCommand();
  execFile(cmd.cmd, cmd.args, function (err, stdout) {
    if (err) return reject({ error: err });
    return resolve(parse(stdout));
  });
}
\end{lstlisting}
\caption{Motivation example of \toolname{}.}
\label{fig:motivation}
\end{figure}

To illustrate this motivation, we conduct a case study on CVE-2020-8132 (Figure~\ref{fig:motivation}) using a vanilla GPT-5-mini agent, comparing its behavior with and without experience guidance. Without experience guidance, the agent exhibits the typical limitations of stochastic iteration: after encountering environment-related obstacles during patching, it repeatedly produces error-prone scripts and gradually shifts from principled repair to superficial modifications aimed only at satisfying test cases. In contrast, when equipped with an experience bank, the agent reuses repair knowledge from an analogous case (CVE-2020-7795) to perform more structured refinement. By retrieving secure repair patterns, such as replacing exec with execFile and adding input validation, it avoids repeated trial-and-error. Moreover, when facing patch conflicts, it follows a more disciplined suspend-diagnose-replan-repair workflow, suggesting that experience guidance improves not only repair correctness but also stability during complex troubleshooting.

\section{Approach}

\subsection{Overview}
\begin{figure*}[htbp]
\centering
\includegraphics[width=1.0\linewidth]{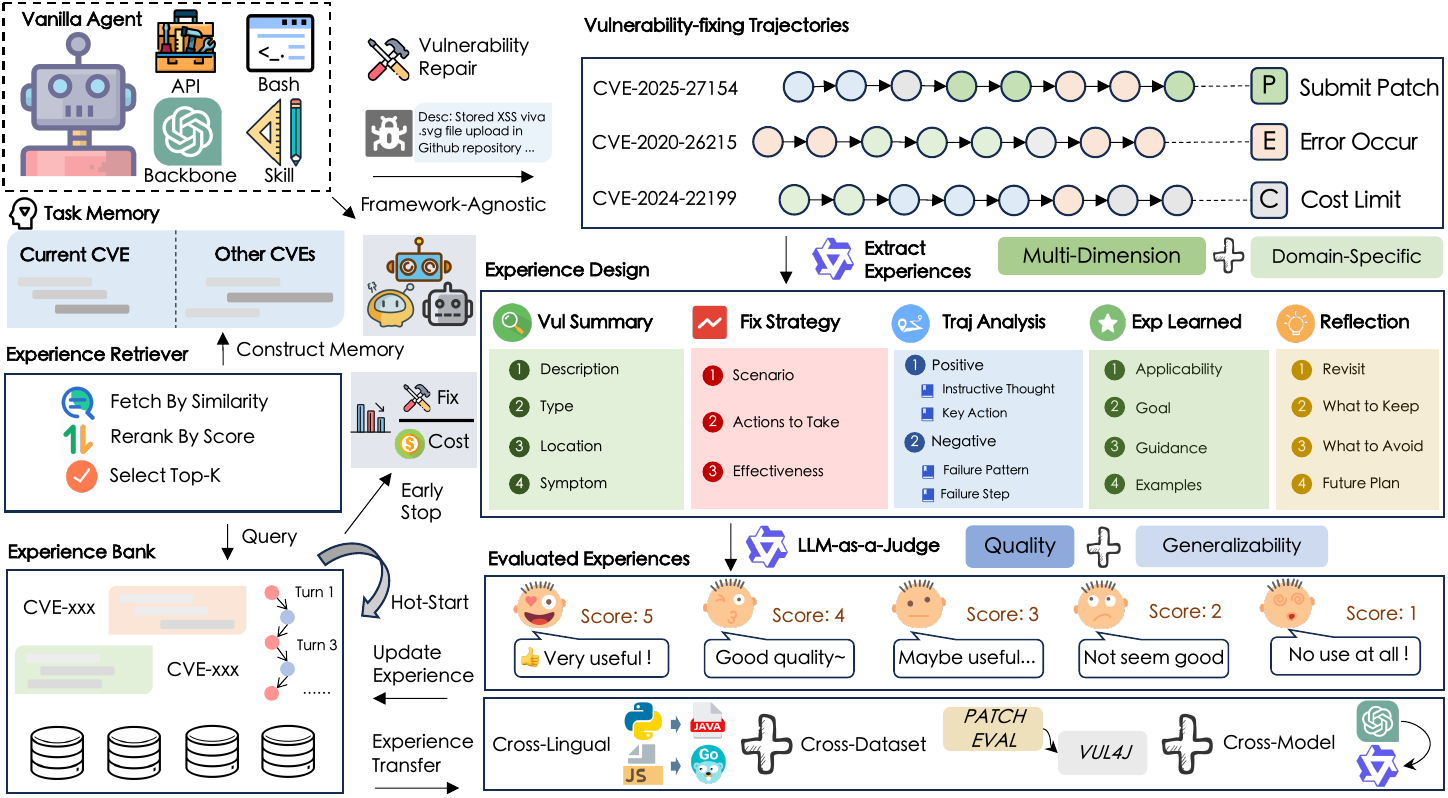}
\caption{Overview of \toolname{}.}
\label{fig:overview}
\end{figure*}

Figure~\ref{fig:overview} presents the overall workflow of \toolname{}. Built on top of a vanilla agent (Figure~\ref{fig:overview}, upper-left), \toolname{} continually accumulates and summarizes domain-specific experience across multi-turn vulnerability repair trajectories, enabling self-evolution in repair performance. In each repair turn, the agent operates on the vulnerabilities that remain unfixed from the previous turn and continues until it either submits a patch or exhausts the repair budget. Below, we briefly describe the main components of \toolname{}.
At the beginning of each turn, \toolname{} retrieves historical experiences relevant to the current vulnerability from the experience bank and injects them into the repair context of the vanilla agent. The agent then interacts with external tools and the target vulnerability's Docker environment to attempt repair within a bounded budget. After the repair process ends, \toolname{} summarizes and reflects on the repair trajectory, compressing it into a reusable domain-specific experience and scoring it based on quality and generality. The resulting experience is stored in the experience bank for future reuse. At the end of each turn, \toolname{} measures the number of newly fixed vulnerabilities and terminates once repair performance converges, at which point it submits all generated patches.
\toolname{} is framework-agnostic and can be adapted to mainstream code-agent frameworks. Moreover, experiences accumulated on one dataset or task can be transferred to other datasets or tasks, supporting cross-language and cross-model transfer. Section~\ref{sec:components} provides detailed descriptions of all components.

\subsection{Definition of Domain-Specific Experience}\label{sec:exp_def}

To make repair experience reusable, retrievable, and transferable across vulnerabilities, we define a standardized domain-specific experience schema with five complementary dimensions. Each dimension is designed to maximize reuse, retrieval precision, and actionable guidance while supporting continuous improvement.
(1) \textit{Vulnerability introduction and analysis} is designed to be a concise, factual description that locates and defines the vulnerability (type, location, reproduction steps, root cause, and affected components). (2) \textit{Repair rationale} is designed to be a short statement of the chosen remediation approach and its justification (method, expected effect, and considered alternatives). (3) \textit{Trajectory analysis} is designed to be a distilled account of the repair process highlighting key successful actions, recurrent failure modes, and representative commands/tests or logs. (4) \textit{Experience summary} is designed to be compact, prescriptive rules derived from the trajectory (preconditions, goal, concrete guidance, and a minimal example). (5) \textit{Reflection \& Improvement} is designed to be a brief evaluation of the outcome, remaining limitations, and suggested follow-ups (e.g., alternative strategies, different repair trajectories). This unified schema enables \toolname{} to continuously accumulate, retrieve, and refine repair knowledge for experience-driven self-evolution.

\subsection{Components}\label{sec:components}

\subsubsection{Experience Retrieval}\label{sec:experience_retrieval}

Before each repair turn, \toolname{} queries the experience bank through an experience retriever to identify relevant prior experiences. For the vulnerability currently under repair, its own historical experiences are always retrieved for direct reuse. For other vulnerabilities, the retriever matches experiences based on CVE and CWE information. Specifically, the retriever first selects the top-$M$ (e.g., 10) candidate experiences from the experience bank according to similarity, as shown in Equation~\ref{eq:rank1}. This step filters out vulnerabilities whose similarity to the current vulnerability is too low, since even high-quality experiences may not be applicable when the underlying vulnerabilities differ substantially. The candidate experiences are then reranked in descending order by a consolidated score $s'$ that combines similarity and the experience score $s_{\mathrm{exp}}$, as shown in Equation~\ref{eq:rank2}. The scoring mechanism for $s_{\mathrm{exp}}$ is described in Section~\ref{sec:exp_construct}. Finally, as shown in Equation~\ref{eq:rank3}, \toolname{} selects the top-$K$ (e.g., 5) experiences with the highest $s'$. This procedure ensures that the retrieved experiences are not only highly similar to the target vulnerability but also sufficiently generalizable to provide useful guidance for the current repair. 

\begin{equation}
\label{eq:rank1}
    \mathcal{C} = \operatorname{TopM}\big(\{\mathrm{sim}(q,e_i)\}_{i=1}^N, M\big)
\end{equation}

\begin{equation}
\label{eq:rank2}
    \mathrm{s'}(e) \;=\; \mu\,\mathrm{sim}(q,e) \;+\; (1-\mu)\,s_{\mathrm{exp}}(e), \quad e\in\mathcal{C}
\end{equation}

\begin{equation}
\label{eq:rank3}
    \mathcal{S} \;=\; \operatorname{TopK}\!\big(\{\mathrm{s'}(e)\}_{e\in\mathcal{C}},\; K\big)
\end{equation}

\noindent\textbf{Data leakage control.} To prevent repair experience from leaking information about the evaluation target, all experience retrieval and cross-dataset transfer settings follow the data-leakage prevention protocol in Section~\ref{sec:leakage_control}.

\subsubsection{Vulnerability Repair}\label{sec:vanilla_agent}

After retrieving relevant experiences, \toolname{} injects them into the repair context to guide the agent in exploring repair trajectories. To minimize the confounding effect of agent-framework design~\cite{hu2026beyond} on the performance of \toolname{}, we construct a general-purpose baseline repair agent, referred to as the \emph{vanilla agent}. The vanilla agent consists of the following components.

\noindent\textbf{Toolkit.} To avoid performance variance introduced by different toolkits, we equip the vanilla agent with only a minimal Bash toolkit that can execute arbitrary shell commands and submit patches via the keyword \texttt{SUBMIT}. We also provide corresponding function-tool API interfaces and adapt them to different model families (e.g., MistralAI).



\noindent\textbf{Memory.} To maintain cross-task knowledge, we equip the vanilla agent with a structured memory module. At the beginning of each repair task, this module is initialized with selected historical experiences retrieved from the experience bank (see Section~\ref{sec:experience_retrieval}). The task memory is organized as a $2 \times 2$ matrix along two dimensions: (1) \emph{Outcome}, which distinguishes successful from failed repair attempts; and (2) \emph{Source}, which distinguishes experiences of the current vulnerability (\emph{self}) from those of other vulnerabilities (\emph{other}). This organization allows the agent to either reuse successful repair patterns or perform negative reasoning to avoid pitfalls observed in previous trajectories.

\noindent\textbf{Repair Context.} In addition to memory, \toolname{} constructs a repair context for the vanilla agent that includes the CVE and CWE information of the target vulnerability, its vulnerable location (if provided), task instructions, and a high-level description of the repair workflow.

\noindent\textbf{Repair Process.} After the repair context is constructed, the vanilla agent initiates the repair process. Following the ReAct~\cite{yao2022react} paradigm, the agent interacts with the Docker environment by invoking the toolkit to autonomously explore repair trajectories. For both cost and efficiency reasons, we limit the number of repair steps and the maximum expenditure. If either limit is exceeded, \toolname{} terminates the repair process early and marks the repair attempt as failed. 

In multi-turn repair, we further define the turn-level yield rate $\alpha$ to measure the cost-effectiveness of the current repair turn, as shown in Equation~\ref{eq:alpha}.

\begin{equation}
\alpha = \frac{\beta}{\gamma}
\label{eq:alpha}
\end{equation}

Here, $\beta$ and $\gamma$ denote the growth rate of repair performance and the growth rate of repair cost, respectively, defined as follows:
\begin{align}
    \beta &= \frac{R_{n} - R_{n-1}}{R_{n-1}} \tag{4.1} \\
    \gamma &= \frac{C_{n} - C_{n-1}}{C_{n-1}} \tag{4.2}
\end{align}

Here, $R_n$ denotes the cumulative successful repair rate after the $n$-th turn, and $C_n$ denotes the cumulative computational cost incurred up to the $n$-th turn. To maintain cost-efficiency under a limited repair budget, we adopt an early-stopping strategy based on a predefined threshold of $\alpha$. When $\alpha$ falls below this threshold, the system terminates the repair process, thereby preventing \toolname{} from continuing ineffective and resource-intensive repair attempts.

\noindent\textbf{Framework-agnostic design.} Although \toolname{} adopts the vanilla agent as the experimental baseline, it can be flexibly adapted to mainstream code-agent frameworks (e.g., SWE-agent~\cite{yang2024swe} and OpenHands~\cite{wang2024openhands}) in practical deployment.

\subsubsection{Experience Construction}\label{sec:exp_construct}

After the vanilla agent completes a repair task, \toolname{} constructs new experiences from the resulting repair trajectory, following the structural specification defined in Section~\ref{sec:exp_def}. Beyond simple experience generation, \toolname{} also refines historical knowledge to support self-evolution. Specifically, by reflecting on the retrieved experiences during analysis of the current trajectory, the agent can synthesize improved experience entries that preserve verified insights while correcting or discarding erroneous patterns. To prevent the context window from becoming a bottleneck over multiple repair turns, we apply a compression mechanism to keep each experience bounded in length while preserving information density. Finally, every synthesized experience is quantitatively evaluated through a scoring process, as described below.

\noindent\textbf{Experience Scoring.} \toolname{} employs an LLM-as-a-Judge strategy to score experiences. As shown in Equation~\ref{eq:score}, the scoring criteria consist of two dimensions: quality and generalizability, where $\lambda$ denotes the weighting coefficient between the two. Quality measures whether an experience can effectively guide future repairs, for example by providing actionable steps or helping avoid incorrect repair paths. Generalizability measures whether the experience can transfer to other vulnerabilities beyond the current one. In our study, we set $\lambda$ to 0.5 to balance quality and generalizability. The judge model assigns high scores only when both dimensions are well satisfied.

\begin{equation}
\label{eq:score}
    s_{\text{exp}} = \lambda \cdot s_{\text{quality}} + (1-\lambda)\cdot s_{\text{general}}
\end{equation}

To mitigate model bias, we experiment with multiple judge models and compare their scores against human ratings. Specifically, we compute the Pearson correlation coefficient (Equation~\ref{eq:pearson}) between model-assigned experience scores and human ratings. Among the evaluated judge models, Qwen3-Max achieves the highest correlation with human judgment. In addition, to reduce variance from a single scoring pass, we score each experience three times and use the mean as the final score.

\begin{equation}
\label{eq:pearson}
    \rho = \frac{\operatorname{cov}(s_{\mathrm{LLM}}, s_{\mathrm{Human}})}{\sigma_{s_{\mathrm{LLM}}}\,\sigma_{s_{\mathrm{Human}}}}
\end{equation}

\subsubsection{Experience Updating}

After an experience is evaluated, \toolname{} persists it together with its score and stores it in the experience bank. For the same vulnerability, \toolname{} compares the newly generated experience $e_t$ from the current turn $t$ with the experience $e_{t-1}$ stored from the previous turn $t-1$, and updates the experience bank accordingly. Specifically, \toolname{} adopts the following update strategies.

\noindent \textbf{Strategy 1:} \textit{Discard}.  
If the score of $e_t$ is lower than that of $e_{t-1}$, \toolname{} discards $e_t$ and retains $e_{t-1}$ as the formal experience entry for turn $t$.

\noindent \textbf{Strategy 2:} \textit{Retain}.  
If the score of $e_t$ is higher than that of $e_{t-1}$, \toolname{} retains $e_t$.

\noindent \textbf{Strategy 3:} \textit{Polish}.  
If the scores of $e_t$ and $e_{t-1}$ are identical, \toolname{} feeds both experiences back into the model and triggers an automated fusion process, in which the model merges complementary details from $e_t$ and $e_{t-1}$ to produce a polished and unified experience entry.

During storage, the experiences are organized as a linked list. For a given vulnerability, experiences generated in different turns are linked sequentially as nodes, and pointers are used to record dependency relations among them. This design allows \toolname{} to access the most recent experiences efficiently while also supporting subsequent offline analysis.

In terms of storage format, \toolname{} maintains both textual and vector representations. It first stores experiences as Markdown files in a hierarchical directory structure, which makes them easy to inspect by researchers and reuse in other repair scenarios. The textual experiences are then vectorized and stored in a vector database to support efficient similarity-based retrieval and reasoning.

After the experience bank is updated, \toolname{} proceeds to the next repair turn and repeats the closed-loop process of experience retrieval, vulnerability repair, experience construction, and experience updating.

\subsubsection{Experience Transfer}

Through the cyclic interaction of the above four components, \toolname{} enables self-evolution of agent-based repair capabilities on arbitrary vulnerability datasets. Moreover, the experiences distilled during this process can be transferred and reused in other vulnerability repair tasks and datasets. This transfer has three key characteristics.
(1) \textbf{Cross-language transfer.} Learned experiences can generalize across programming languages (e.g., Python $\rightarrow$ Java) without manual adaptation, demonstrating the language-agnostic nature of our approach.
(2) \textbf{Cross-dataset / cross-task transfer.} Experiences accumulated on one dataset or task can be reused effectively on unseen datasets and new repair scenarios, thereby avoiding repeated learning.
(3) \textbf{Cross-model transfer.} The acquired experiences are independent of any specific model architecture and can be transferred across different backbone models, improving the flexibility and applicability of \toolname{}.

\section{Experimental Setup}
\subsection{Research Questions}
We evaluate \toolname{} on the following research questions:

\noindent\textbf{RQ1:} How does \toolname{} compare against existing vulnerability repair methods?

\noindent\textbf{RQ2:} How does \toolname{} improve repair performance through self-evolution over multiple repair turns?

\noindent\textbf{RQ3:} How effective is each component of \toolname{}?

\noindent\textbf{RQ4:} Can the experience generated by \toolname{} be transferred across languages and datasets?

\noindent\textbf{RQ5:} How does \toolname{} balance cost and performance?

\subsection{Datasets}
We evaluate \toolname{} on the latest multilingual (JavaScript, Python, and Go) benchmark PATCHEVAL~\cite{wei2025patcheval} and C-based benchmark SEC-bench~\cite{lee2025sec}. 
PATCHEVAL comprises 1,000 vulnerabilities (230 w/ docker + 770 w/o docker), we use a subset of 230 instances with accompanying Docker images to evaluate \toolname{}. 
SEC-bench comprises 200 C-language instances across 29 projects, including the corresponding vulnerability descriptions, sanitizer reports, and CVE identifiers. 
Following the original papers, we use two complementary repair settings:
oracle localization for PATCHEVAL and end-to-end repair for SEC-bench.
In addition, we employ VUL4J~\cite{bui2022vul4j} to further validate the experience transfer capability of \toolname{}. VUL4J consists of 79 Java CVEs, which is orthogonal to both PATCHEVAL and SEC-bench in terms of programming language and CVE composition. Therefore, it serves as a suitable and independent benchmark for sufficiently verifying the transfer ability of \toolname{}.

\subsection{Baselines and Models}
On PATCHEVAL and SEC-bench, we evaluate \toolname{} against 15 baselines: 2 learning-based methods (VulRepair~\cite{fu2022vulrepair}, VulMaster~\cite{zhou2024out}), 6 LLM-based methods (Zero-shot~\cite{kojima2022large}, Few-shot~\cite{brown2020language}, ChatRepair~\cite{xia2024automated}, PailGen~\cite{shao2026fix}, LoopRepair~\cite{ye2026well}, IntentFix~\cite{jin2026intent}), and 7 agent-based methods (SmolAgent~\cite{roucher2025smolagents}, AgentMem~\cite{xu2025amem}, OpenHands~\cite{wang2024openhands}, Claude Code~\cite{anthropic2025claudecode}, Vanilla Agent, Live-SWE-Agent~\cite{xia2025live}, ContraFix~\cite{liu2026contrafix}). We prioritize using publicly available results from previous work; for results that are not available, we reproduce them using the original paper’s code configuration.
Specifically, we adopt the official SEC-bench results for SmolAgent and AgentMem, and include the OpenHands, Claude Code, and ContraFix results reported by ContraFix~\cite{liu2026contrafix}. Since OpenHands and Claude Code use the stronger GPT-5 backbone, their comparison with \toolname{} using GPT-5-mini is conservative. We reproduce VulRepair and VulMaster with CodeT5. For the remaining baselines, we use GPT-5-mini as a unified base model to ensure a fair comparison in the main experiment. In addition, we study five representative backbone LLMs, including three closed-source models (GPT-5-mini~\cite{singh2025openai}, DeepSeek-v3.1~\cite{liu2024deepseek}, and Qwen3.5-Plus~\cite{yang2025qwen3}) and two open-source models (Devstral-24B~\cite{rastogi2025devstral} and Devstral-123B~\cite{rastogi2025devstral}). On VUL4J, we compare \toolname{} with 5 advanced vulnerability repair baselines~\cite{xia2024automated, bui2024apr4vul, ntr, vrpilot, hu2025tsapr}.

\subsection{Evaluation Metrics}
Following prior studies~\cite{lee2025sec,zhang2026vulnresolver}, we use two metrics to evaluate the effectiveness of \toolname{} under the official benchmark repair oracle. \textit{\#Fix} measures the number of vulnerabilities whose generated patches pass the benchmark-provided evaluation scripts. \textit{\%Fix} measures the corresponding proportion among all evaluated vulnerabilities. These scripts execute the associated PoC tests and unit tests, and the same oracle is applied to \toolname{} and all baselines to ensure fair relative comparison.

\subsection{Implementation Details}
To implement \toolname{}, we first build a basic, general-purpose vulnerability‑repair agent, which we refer to as the vanilla agent. The vanilla agent follows the widely used Mini‑SWE‑Agent paradigm~\cite{yang2024swe}; we further adapt it for vulnerability repair and extend the existing bash tool with a function‑tool interface to support various LLMs. Building on this foundation, we implement the components of \toolname{}. Specifically, when constructing the experience bank we persist experiences as markdown using templates and vectorize them into Chroma. For experience retrieval, we use the LangChain framework with bge‑large‑en‑v1.5 as the embedding model for similarity matching. For experience extraction and evaluation, we choose Qwen3‑Max for its low cost and strong overall capabilities. When evaluating API models, we use OpenAI’s official API for the GPT family, Bailian's DeepSeek API for the DeepSeek models, and deploy other open‑source models with vLLM on an 8× RTX 5880 server using bf16 precision and a context length limit of 64,000 tokens. For reproducibility, all models are generated with temperature set to 0. \toolname{} performs continuous multi-turn repairs on unresolved vulnerabilities, with a maximum of 15 turns in this work. If there are no newly fixed vulnerabilities in the current turn, we consider the agent to have reached its repair capacity and terminate the evaluation. We align the evaluation budget and stopping conditions across methods where applicable, while preserving each framework's native interaction style and tooling assumptions.
For each vulnerability, we cap \toolname{}'s repair effort at 100 steps or \$3, and the repair process terminates early if either condition is met. All experiments were run on a Linux server with 740 GB of RAM. We set the number of concurrent threads to 32 to speed up the experiments.

\section{Evaluation and Results}
\subsection{RQ1: Overall Performance}


Table~\ref{tab:vul_repair_main} shows that \toolname{} achieves the best overall performance on both PATCHEVAL and SEC-bench. Specifically, \toolname{} repairs 215/230 vulnerabilities on PATCHEVAL (93.47\%) and 174/200 on SEC-bench (87.00\%), yielding an overall repair rate of 90.46\% over 430 instances. Compared with existing state-of-the-art repair methods, \toolname{} maintains a clear advantage. Among prior approaches with full results on our 430-instance setting, the strongest result is achieved by the recent self-evolving SE agent Live-SWE-Agent, which attains an overall repair rate of 83.48\%. ContraFix achieves 72.17\% on PATCHEVAL, 84.00\% on SEC-bench, and 77.67\% overall under the same 430-instance normalization. Notably, while both Live-SWE-Agent and \toolname{} are built upon a self-evolving agent framework, \toolname{} achieves a further 6.98\% improvement overall, highlighting the advantage of experience-driven evolution over tool-centric self-evolution in vulnerability repair.
In contrast, earlier LLM-based methods such as ChatRepair and LoopRepair reach only 50.69\% and 53.72\%, while learning-based methods perform substantially worse, likely because approaches such as VulRepair and VulMaster are constrained by fixed input formats and language-specific settings (e.g., C), and are primarily designed for relatively short repair contexts. These results show that our method substantially enhances the effectiveness of LLM-based and agent-based vulnerability repair, surpassing not only conventional repair paradigms but also prior self-evolving agent techniques.

\input{tab/overall_performance}

\subsection{RQ2: Turn-level Performance}
\begin{figure*}[t]
    \centering
    \makebox[\textwidth][c]{%
    \subfigure[GPT-5-mini]{%
        \includegraphics[width=0.19\textwidth]{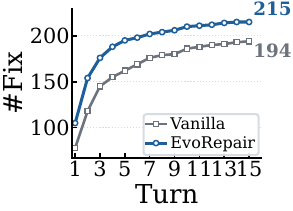}
        \label{fig:tts_gpt}
    }\hfill
    \subfigure[Qwen3.5-Plus]{%
        \includegraphics[width=0.19\textwidth]{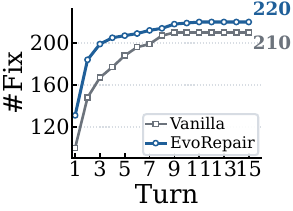}
        \label{fig:tts_qwen}
    }\hfill
    \subfigure[DeepSeek-v3.1]{%
        \includegraphics[width=0.19\textwidth]{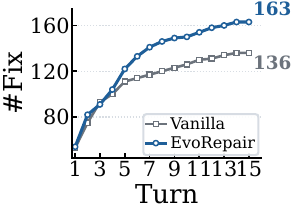}
        \label{fig:tts_ds}
    }\hfill
    \subfigure[Devstral-24B]{%
        \includegraphics[width=0.19\textwidth]{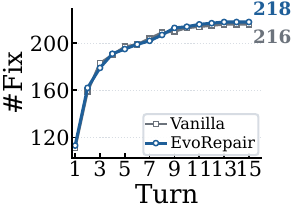}
        \label{fig:tts_dev_24b}
    }\hfill
    \subfigure[Devstral-123B]{%
        \includegraphics[width=0.19\textwidth]{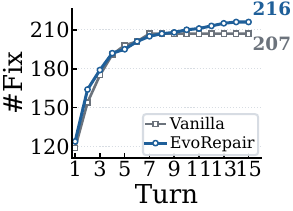}
        \label{fig:tts_dev_120b}
    }%
    }
    \caption{Turn-level performance on PATCHEVAL.}
    \label{fig:tts}
\end{figure*}

\begin{figure}[htbp]
    \centering
    \subfigure[GPT-5-mini]{%
        \includegraphics[width=0.333\columnwidth]{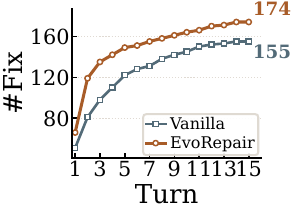}%
        \label{fig:tts_gpt_sec}%
    }\hfill
    \subfigure[Qwen3.5-Plus]{%
        \includegraphics[width=0.333\columnwidth]{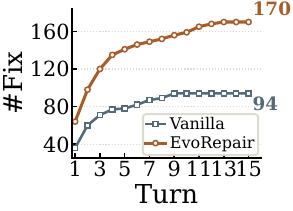}%
        \label{fig:tts_qwen_sec}%
    }\hfill
    \subfigure[DeepSeek-v3.1]{%
        \includegraphics[width=0.333\columnwidth]{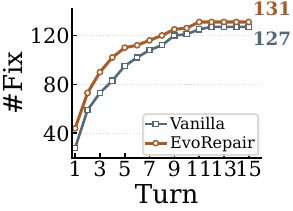}%
        \label{fig:tts_ds_sec}%
    }
    \caption{Turn-level performance on SEC-bench.}
    \label{fig:tts_sec}
\end{figure}

Figure~\ref{fig:tts} and Figure~\ref{fig:tts_sec} show the turn-level performance of \toolname{} across 15 repair turns on PATCHEVAL and SEC-bench. We summarize three key findings:
(1) \textbf{Breaking the Performance Ceiling}.
Across both datasets, \toolname{} consistently fixes more vulnerabilities than the vanilla agent. On PATCHEVAL, final fixes improve from 194 to 215 on GPT-5-mini, 210 to 220 on Qwen3.5-Plus, and 136 to 163 on DeepSeek-v3.1; on SEC-bench, they improve from 155 to 174, 94 to 170, and 127 to 131, respectively.
(2) \textbf{Stable Gains and Faster Convergence}.
\toolname{} stays above the vanilla agent for most repair turns and usually reaches near-peak performance within the first several turns, especially on GPT-5-mini and Qwen3.5-Plus. This indicates that self-evolution reduces redundant trial-and-error during iterative repair.
(3) \textbf{Robustness Across Model Families and Scales}.
The gains hold across proprietary and open-source backbones, including GPT-5-mini, Qwen3.5-Plus, DeepSeek-v3.1, Devstral-24B, and Devstral-123B, suggesting that experience-based self-evolution generalizes across model families and scales.

\subsection{RQ3: Ablation Study}

In RQ3, we conduct ablation studies on the core components and configurations of \toolname{}, including the number of retrieved experiences, the mechanism for experience retrieval, the prompting strategies for experience construction, and the methodologies employed to address the cold-start problem of the experience bank. We choose PATCHEVAL for ablation study because it is multilingual and larger in scale compared to SEC-bench.

\begin{figure*}[htbp]
\centering
\includegraphics[width=0.9\linewidth]{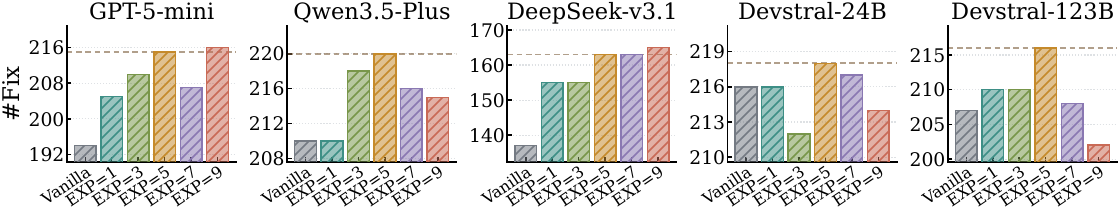}
\caption{Relationship between number of retrieved experiences and repair performance of \toolname{}.}
\label{fig:ablation_1}
\end{figure*}

\noindent\textbf{Ablation 1: Experience Count.}
When constructing the repair context, \toolname{} retrieves a specific number of experiences from the experience bank to serve as guidance. To investigate the relationship between the experience count and the resulting repair effectiveness, we conduct experiments by retrieving $k$ experiences, where $k \in \{1, 3, 5, 7, 9\}$.
As illustrated in Figure~\ref{fig:ablation_1}, we evaluate the repair efficacy of \toolname{} across varying quantities of retrieved experiences. Overall, the repair performance exhibits a characteristic bell-shaped trajectory relative to the experience count, initially ascending to a peak before undergoing a marginal decline. Empirical evidence suggests that Exp=5 represents the optimal configuration for the majority of the evaluated models; specifically, it yields the highest repair count in three out of the five models and secures the second-best performance in the other two.
\textbf{Consequently, we adopt the Exp=5 configuration for all subsequent experiments.}

Furthermore, the learning capacity for retrieved experiences varies across different model architectures. SOTA commercial models, such as Qwen3.5-Plus and GPT-5-mini, demonstrate a stable performance gain as the experience count increases. Even when additional experiences introduce potential noise, these models exhibit only marginal performance degradation, underscoring their superior robustness. In contrast, while the older DeepSeek-v3.1 benefits significantly from a larger context (peaking at Exp=9), indicating commendable In-Context Learning (ICL) capabilities, it remains constrained by lower performance bounds and higher sensitivity to the quantity of guidance. This suggests a persistent gap in reasoning maturity compared to the latest LLMs.

For code-specialized models like Devstral, although they possess high intrinsic repair proficiency, they derive limited incremental utility from retrieved experiences. In fact, excessive experiences often lead to diminishing returns or negative performance impacts. This phenomenon suggests that while code-specific models excel in structural syntax and local logic, their ICL efficacy is heavily dictated by their underlying natural language understanding and knowledge transfer abilities. Consequently, the high specialization of code models may lead to a "knowledge saturation" point where extra context acts more as a distractor than a catalyst, highlighting a critical trade-off between domain expertise and contextual adaptability.

\input{tab/ablation_score} 
\noindent\textbf{Ablation 2: Experience Retrieval Mechanism.}
When retrieving experiences, \toolname{} simultaneously considers the similarity between vulnerabilities and the score (quality+generalizability) of the experiences. To investigate the impact of experience score on retrieval effectiveness, in Ablation 2, we exclude the experience scoring component and retrieve experiences based solely on vulnerability similarity.
As shown in Table~\ref{tab:ablation_score}, we find that the removal of the scoring component leads to a consistent performance degradation across all models to varying degrees. This observation suggests that vulnerability similarity alone is insufficient to guarantee the efficacy of retrieved guidance. The results underscore the critical role of the scoring module in filtering out low-quality or noisy experiences, thereby refining the retrieval results beyond raw similarity to provide more robust guidance for the repair process.

\noindent\textbf{Ablation 3: Experience Construction Strategy.} To ensure that the experiences generated by \toolname{} have good generalization while being easily practical for agents, we adopt an example-driven strategy in constructing experiences. This involves requiring the model to provide minimal examples, including script commands and expected outcomes, while summarizing the experiences in natural language. To verify the effectiveness of the example-driven strategy, we remove this approach in Ablation 3, requiring the model only to summarize experiences in natural language for comparison of repair results.
As shown in Table~\ref{tab:ablation_example_driven}, all models achieve performance gains when employing the example-driven strategy, indicating that concrete examples effectively assist models in better internalizing and leveraging learned experiences.
\begin{table}[htbp]
\centering
\caption{Effectiveness of example-driven experience construction.}
\label{tab:ablation_example_driven}
\begin{tabular}{lrr}
\toprule
Model & w/o example & w/ example \\
\midrule
GPT-5-mini & 209 & 215 (\textcolor{deepgreen}{+6}) \\
DeepSeek-v3.1 & 160 & 163 (\textcolor{deepgreen}{+3}) \\
Qwen3.5-Plus & 207 & 220 (\textcolor{deepgreen}{+13}) \\
Devstral-24B & 214 & 218 (\textcolor{deepgreen}{+4}) \\
Devstral-123B & 207 & 216 (\textcolor{deepgreen}{+9}) \\
\bottomrule
\end{tabular}
\end{table}

\noindent\textbf{Ablation 4: Cold-start Strategy.}
To analyze cold-start behavior, we compare two strategies in the first repair turn: (1) \textbf{Pre-repair}, which provides pre-generated experiences from other vulnerabilities; and (2) \textbf{Standard Patch}, which provides standard patches from 770 vulnerabilities outside the PATCHEVAL evaluation set as few-shot guidance. Both settings follow the data-leakage prevention protocol in Section~\ref{sec:leakage_control}. Table~\ref{tab:ablation_cold_start} reports the number of vulnerabilities fixed in the first turn, with parenthesized values denoting changes relative to the setting without cold-start guidance. Under this first-turn setting, Pre-repair consistently improves over the no-guidance setting across all models, whereas directly retrieving similar vulnerability patches as few-shot examples can be unstable and even leads to a substantial performance drop (-21) on GPT-5-mini. These results suggest that trajectory-derived experiences provide more effective first-turn guidance than raw patch examples in the cold-start setting.
\input{tab/ablation_cold_start}


\subsection{RQ4: Experience Transfer Performance}
We select Qwen3.5-Plus, which achieves the best performance on PATCHEVAL, as the teacher model to transfer its synthesized experiences to VUL4J. The transferred experience bank is frozen before VUL4J evaluation, and VUL4J repair trajectories do not update or rescore it, following Section~\ref{sec:leakage_control}. For the student model, we employ two configurations: in configuration T1, the student model is also Qwen3.5-Plus, intended to validate intra-model experience transfer; in configuration T2, the student model is Qwen3-Max, utilized to evaluate cross-model transferability. As shown in Table~\ref{tab:vul}, under both configurations, EvoRepair outperforms all baselines, achieving performance improvements of 9.67\% and 8.69\%, respectively, compared to repair without experience transfer. The transfer results suggest that some experiences learned on one benchmark can remain useful across datasets and programming languages.

\input{tab/vul4j}

\subsection{RQ5: Cost Analysis}
As shown in Table~\ref{tab:cost_analysis}, \toolname{} keeps the total cost nearly unchanged compared with the vanilla agent (\$1,663.7 vs. \$1,666.9, -0.19\%). The cost impact varies by backbone: DeepSeek-v3.1 increases most (+41.86\%) because many unrecoverable cases hit the per-instance cost limit, while Qwen3.5-Plus reduces cost by 41.58\% as experience guidance accelerates convergence. Other models increase by roughly 10\%--20\%, mainly from the prompt-token overhead of injected experiences.
\input{tab/cost}
Trajectory overhead is also modest: \toolname{} shortens average repair paths on three of five models, with the largest reduction on Qwen3.5-Plus (-14.3\%), and reduces total repair turns for two models. These results suggest that early stopping and experience compression are promising cost-control directions for self-evolving AVR.
\input{tab/early_stop_summary}

\textbf{Early Stop.} In Table~\ref{tab:early_stop_summary}, we discuss the efficacy of the threshold-based early-stopping strategy. The turn-level yield rate $\alpha$, as defined in Equation~\ref{eq:alpha}, represents the ratio between the marginal gain in repair effectiveness and the marginal increase in repair cost. The results demonstrate that by setting the threshold between 0.15 and 0.25, all models can significantly reduce computational costs ($\downarrow$ 20\%-40\%) with minimal degradation in performance ($\downarrow$ 1\%-10\%), thereby validating the effectiveness of the early-stopping strategy.

\section{Discussion}
\subsection{Test-Time Learning Protocol and Data-Leakage Prevention}\label{sec:leakage_control}

\toolname{} is evaluated under an online, chronological self-evolving test-time learning protocol rather than a conventional one-shot, no-memory protocol. 
Under this protocol, the agent may learn only from interaction signals produced during its own repair attempts and from experiences generated by vulnerabilities available earlier in time; such experiences are the intended non-parametric knowledge accumulated during deployment, rather than external supervision or leaked ground truth. 
This setting follows the test-time adaptation practice used in prior agent work such as Reflexion~\cite{shinn2023reflexion} and CLIN~\cite{majumder2023clin}, where agents repeatedly interact with an environment, summarize failed or successful trajectories into textual memory, and reuse that memory in subsequent trials.

We distinguish intended self-evolution from data leakage as follows. Data leakage occurs when the repair of a target CVE can access privileged information that would not be available at deployment time, including the target CVE's reference patch, hidden evaluation labels, duplicate vulnerabilities, and later-disclosed CVEs. 
In contrast, self-evolving test-time learning only uses non-parametric textual experiences generated from the agent's own previous interactions, executable feedback available during repair, and earlier-disclosed vulnerabilities.
To enforce this boundary, we process CVEs in chronological order according to their disclosure dates and restrict cross-vulnerability retrieval to earlier-disclosed CVEs within the same benchmark.
During iterative repair of the same CVE, only experiences generated in previous turns of that CVE are reused, analogous to trial-level memory in test-time agent adaptation.
For cross-dataset transfer experiments, the transferred experience bank is constructed before evaluating the target dataset and then frozen.

To assess whether processing order changes experience availability, we examine when the bank is updated. Each turn reads the bank frozen at the end of the preceding turn, so within-turn order does not affect available experiences. To clarify the role of external initialization, Pre-repair draws on the 770 PATCHEVAL patches without Docker environments and is used only in the cold-start ablation, separately from the main evaluation.

\subsection{Experience Sources and Attribution}
To examine whether additional fixes rely exclusively on successful, same-CVE, or same-CWE experience, we inspect all experiences injected into the final successful turns of PATCHEVAL vulnerabilities repaired by \toolname{} but not vanilla (Table~\ref{tab:experience_sources}). Failed attempts supply 27.6\% of these experiences, other CVEs supply 82.2\%, and different CWEs supply 30.1\%. These proportions describe usage rather than independently establish each source's causal contribution.

\begin{table}[htbp]
\centering
\caption{Experience sources in final successful turns of PATCHEVAL cases repaired only by \toolname{}. Each pair partitions all injected experiences.}
\label{tab:experience_sources}
\small
\begin{tabular}{llr}
\toprule
Dimension & Source & Share \\
\midrule
Outcome & Successful attempts & 72.4\% \\
        & Failed attempts & 27.6\% \\
CVE & Same CVE & 17.8\% \\
    & Other CVEs & 82.2\% \\
CWE & Same CWE & 69.9\% \\
    & Different CWEs & 30.1\% \\
\bottomrule
\end{tabular}
\end{table}

\subsection{Human Validation of Experience Scoring}
\begingroup
\setlength{\columnsep}{6pt}
\setlength{\intextsep}{4pt}
\begin{wraptable}{r}{0.52\columnwidth}
\centering
\caption{Human and model score counts for 100 experiences.}
\label{tab:judge_validation}
\small
\begin{tabular}{ccc}
\toprule
Score & Human & Qwen3-Max \\
\midrule
1 & 3 & 0 \\
2 & 18 & 18 \\
3 & 24 & 21 \\
4 & 45 & 61 \\
5 & 10 & 0 \\
\bottomrule
\end{tabular}
\end{wraptable}
Because judge errors can affect retrieval and updating, we assess agreement with human judgment using 50 successful and 50 failed experiences. Three annotators independently rate each on a five-point scale; majority ratings are used, with three-way disagreements resolved through discussion. Table~\ref{tab:judge_validation} reports the score distributions. Against adjudicated human ratings, Qwen3-Max achieves Pearson $r=0.7313$, Spearman $\rho=0.7573$, 73\% exact agreement, 97\% agreement within one point, MAE of 0.320, and RMSE of 0.678. The model favors intermediate scores and assigns neither extreme. This agreement supports consistency with human judgment, but does not establish whether scores predict repair gains on unseen CVEs.

To check sensitivity to the quality--generalizability trade-off, we conduct a preliminary analysis of $\lambda$. Results suggest stability at $\lambda=0.5$; broader sensitivity remains unquantified.
\par
\endgroup

\subsection{Cost Accounting and Comparison}
To assess efficiency against another self-evolving agent, we compare cumulative costs with Live-SWE-Agent~\cite{xia2025live}, including failed attempts. A step is one interaction within a trajectory; a turn is a repair round followed by experience summarization and updating. Mean trajectory length includes successful and failed attempts, and total cost accumulates all attempts across all turns. Complementing Table~\ref{tab:cost_analysis}, Live-SWE-Agent with GPT-5-mini on PATCHEVAL averages 20.81 steps, 3.47 turns, 317.3K prompt tokens, and 35.7K response tokens per CVE, at a cumulative cost of \$186.39. Thus, \toolname{} repairs 8.0\% more cases at 4.30\% greater cost. The near-zero aggregate cost change across backbones reflects offsetting increases and decreases, rather than uniform cost neutrality.

\subsection{Threats to Validity}

\textbf{Internal Validity.} Internal validity concerns potential experimental biases that may affect the fairness or consistency of our evaluation. To reduce the impact of randomness during the generation process of LLMs on the experimental results, we repeat each experimental setting three times in the same hardware environment and take the median. To quantify run-to-run variability, we also record the standard deviation in repaired-case counts, which is below 0.5 across models. Additionally, we save the trajectories generated during the agent's repair process and conduct offline analysis and comparison of multiple trajectories for the same CVE to validate the consistency of the agent's behavior. All experiments are conducted in isolated Docker environments to ensure that they are not affected by other processes.

\noindent\textbf{External Validity.} External validity addresses the generalizability of our findings beyond the specific datasets and programming languages used in this study. Regarding the dataset issue, to validate the effectiveness of \toolname{} across different datasets, we conduct extensive experiments on three high-quality vulnerability datasets (PATCHEVAL, SEC-bench, VUL4J) with five open-source and closed-source models. The three datasets cover vulnerabilities in five different programming languages (C, Java, Python, JavaScript, Go), demonstrating that the effectiveness of \toolname{} is independent of the programming language.
To evaluate C repair at a larger scale, we selected the newer, larger SEC-bench over San2Vuln (27 C/C++ cases). Generalization under rapid distribution shifts and integration into SWE-agent or OpenHands remain untested; framework independence is a design claim.

\section{Conclusion}
This paper presents \toolname{}, an experience-based self-evolving AVR framework that addresses the lack of intra-vulnerability experience accumulation and cross-vulnerability experience reuse. Evaluations on PATCHEVAL, SEC-bench, and VUL4J show that \toolname{} improves over learning-based, LLM-based, and agent-based baselines, while ablations confirm the benefits of experience retrieval, quality-aware scoring, and cold-start design. Beyond aggregate repair rates, these results suggest that compact, quality-aware experience can preserve security reasoning across attempts, help agents avoid repeated failures, and improve repair stability.

\section*{Acknowledgments}
The authors would like to thank the anonymous reviewers for their insightful comments.
This research was supported by the Natural Science Foundation of Jiangsu Province (BK20251458), and the National Natural Science Foundation of China (No.62471235, No.U24A20337, No.62502211).

\section*{Data Availability}
The artifact, including code, scripts, data and results, is available at https://github.com/iSEngLab/Learn-from-Repair-Experience.
\bibliographystyle{IEEEtran}
\bibliography{reference}

\newpage
\appendix

\section{Algorithm}
As shown in Figure~\ref{alg:self_evolving_framework}, we provide a complete pseudocode representation of the \toolname{} workflow, which delineates the step-by-step logic and decision-making mechanisms integrated into each remediation cycle.
\input{tab/algorithm}

\section{Full Result on SEC-bench}
We present in Table~\ref{tab:tts_sec} the complete results of 15 turns of repairs on SEC-bench for the Vanilla agent, Live-SWE-Agent, and \toolname{}. The results show that \toolname{} generally outperforms both Live-SWE-Agent and the Vanilla agent. When using Qwen3.5-Plus as the base model, \toolname{} achieves the most pronounced improvement, fixing 26 more vulnerabilities than Live-SWE-Agent and 76 more than the Vanilla agent. With DeepSeek-v3.2 and GPT-5-mini, although \toolname{}’s final repair ceiling is roughly on par with Live-SWE-Agent, \toolname{} typically converges 3–5 turns faster. Additionally, in Tables~\ref{tab:sec_file_loc} and ~\ref{tab:sec_method_loc} we report each method’s file-level and function-level localization accuracy at every turn.
\input{tab/secbench_tts}
\input{tab/secbench_file_loc}
\input{tab/secbench_method_loc}

\section{Overlap Analysis}
\begin{figure}[htbp]
     \centering
     \vspace{-2em}
     \subfigure[GPT-5-mini]{
        \includegraphics[width=0.4\columnwidth]{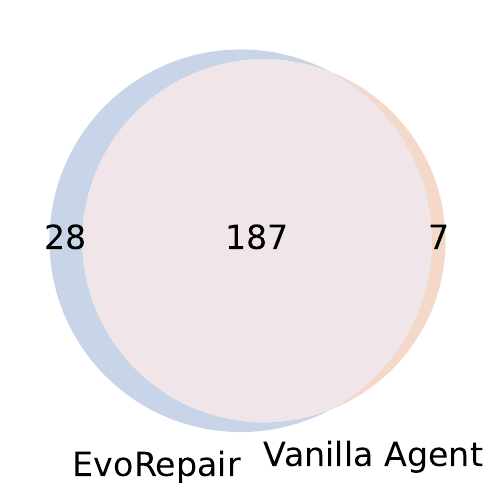}
        \label{fig:overlap_gpt}
    }
    \subfigure[Qwen3.5-Plus]{
        \includegraphics[width=0.4\columnwidth]{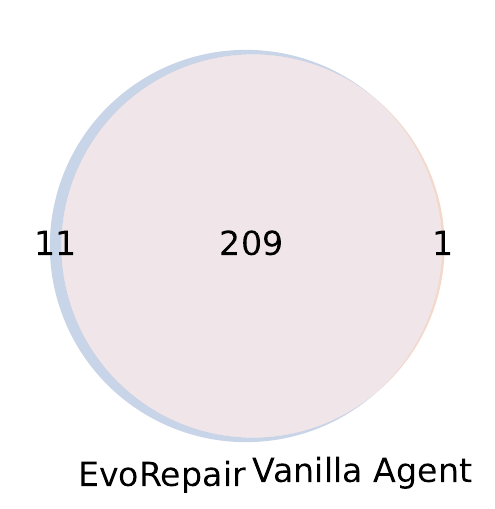}
        \label{fig:overlap_qwen}
    }
    \subfigure[DeepSeek-v3.1]{
        \includegraphics[width=0.4\columnwidth]{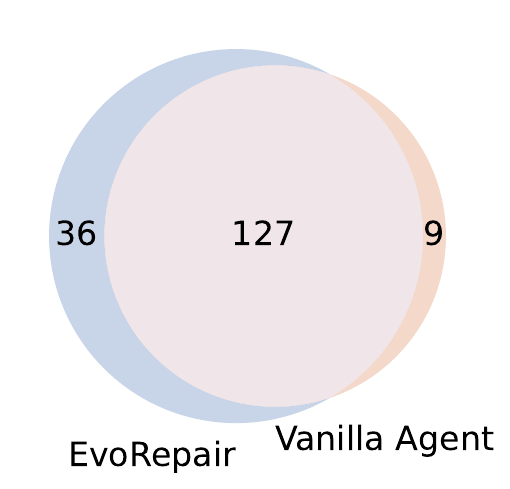}
        \label{fig:overlap_ds}
    }
    \subfigure[Devstral-24B]{
        \includegraphics[width=0.4\columnwidth]{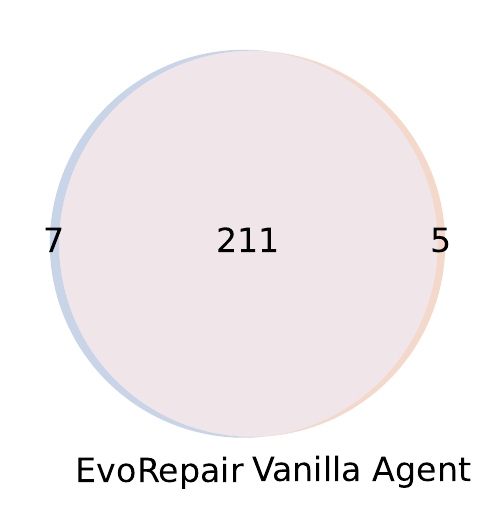}
        \label{fig:overlap_dev_24b}
    }
    \subfigure[Devstral-123B]{
        \includegraphics[width=0.4\columnwidth]{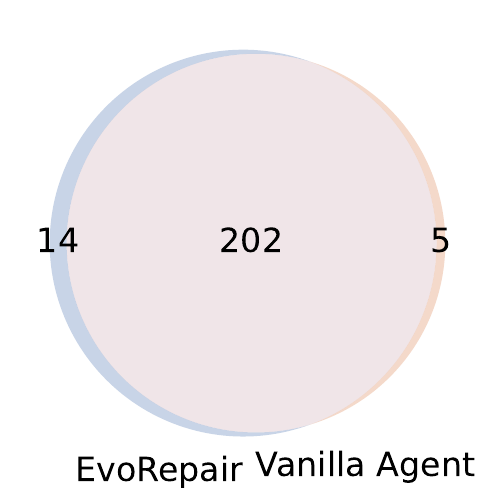}
        \label{fig:overlap_dev_120b}
    }
     \label{fig:overlap_method}
\end{figure}

\textbf{Overlap analysis between vanilla agent and \toolname{}}. on PATCHEVAL, we conduct an overlap analysis of the vulnerabilities remediated by \toolname{} and the vanilla agent to evaluate their respective repair capabilities. As illustrated in Figures~\ref{fig:overlap_gpt} to~\ref{fig:overlap_dev_120b}, we compare the CVEs remediated by \toolname{} and the vanilla agent across various base models. We observe a high degree of overlap between the vulnerabilities resolved by the two methods. Notably, \toolname{} effectively encompasses the remediation coverage of the vanilla agent; specifically, when using Qwen3.5-Plus as the base model, there is only one vulnerability that \toolname{} fails to remediate while the vanilla agent succeeds. This indicates that \toolname{} successfully preserves the baseline agent's strengths while simultaneously augmenting its repair capabilities. These results demonstrate the robustness of \toolname{}, confirming that the integration of the experience library does not introduce excessive noise or degrade the original repair performance of the base model.

\begin{figure}[htbp]
     \centering
     \subfigure[Vanilla Agent]{
        \includegraphics[width=0.46\columnwidth]{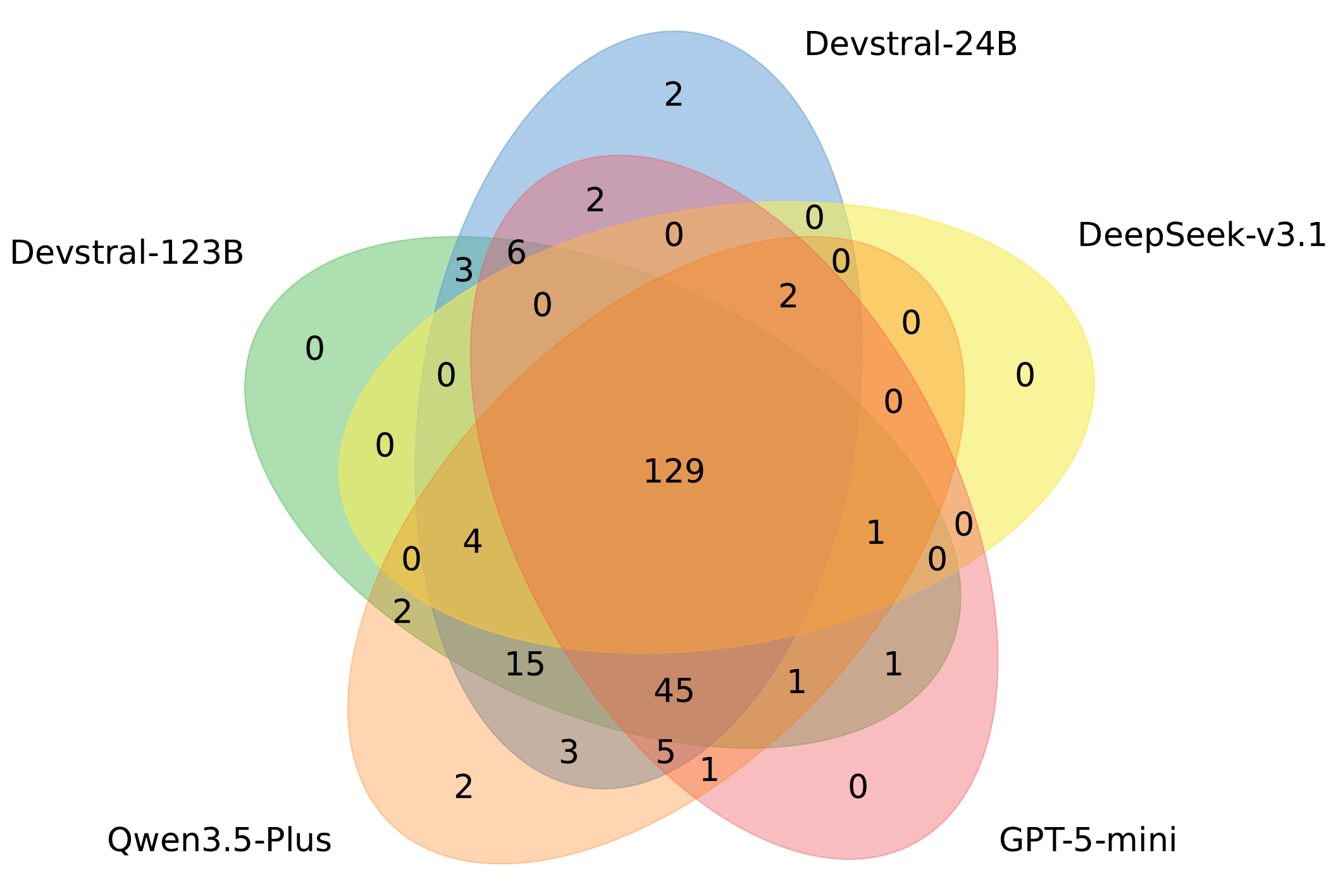}
        \label{fig:overlap_vanilla}
    }
    \subfigure[\toolname{}]{
        \includegraphics[width=0.46\columnwidth]{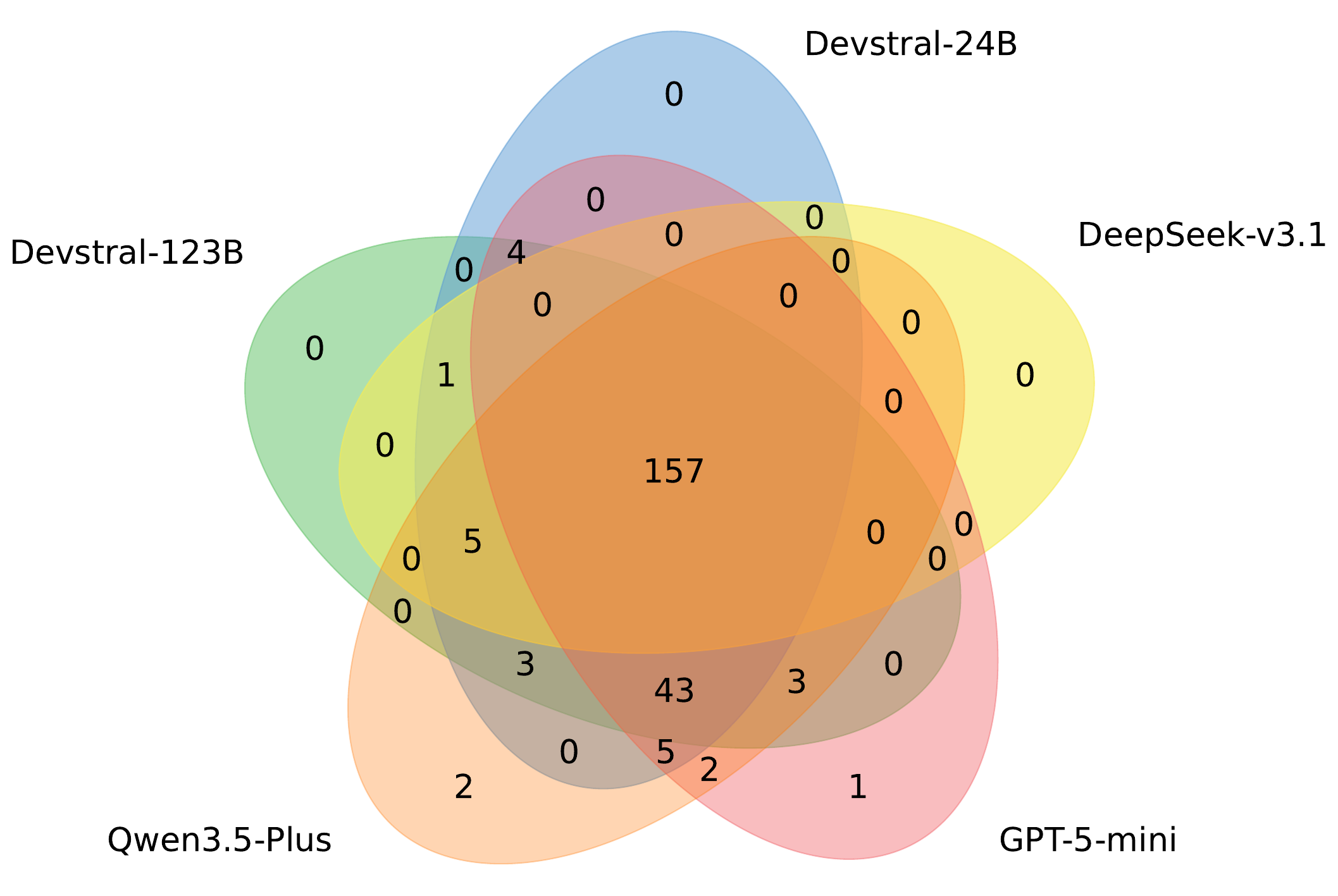}
        \label{fig:overlap_evorepair}
    }

     \label{fig:overlap_model}
\end{figure}

\textbf{Overlap analysis between different models}. Beyond analyzing the overlap in remediated vulnerabilities across different methods, we also investigate the overlap among various base models. As illustrated in Figure~\ref{fig:overlap_vanilla}, we observe that even prior to the integration of \toolname{}, the vanilla agents across different base models already exhibit a significant degree of overlap in their remediation coverage. Specifically, the 129 jointly repaired vulnerabilities account for 60\% or more of the successful repairs for any individual model, indicating a substantial level of inherent consistency in how different LLMs handle standard security flaws.

However, upon the introduction of \toolname{}, as shown in Figure~\ref{fig:overlap_evorepair}, the number of shared successful repairs across the five models increases to a striking 157, representing over 70\% of any single model's repair set. This significant expansion in the overlap set demonstrates that \toolname{} successfully preserves the original strengths of the vanilla agents while consistently augmenting their capabilities. More importantly, this trend suggests that the experience library acts as a powerful 'knowledge anchor,' harmonizing the repair trajectories of diverse base models toward a convergent, optimal solution space. It further proves that \toolname{} possesses excellent robustness, as the high-quality diagnostic priors effectively guide different models to resolve complex vulnerabilities without introducing excessive noise that might otherwise degrade their baseline performance.

\section{Combined With Live-SWE-Agent}
During the reproduction of Live-SWE-Agent, we observe that while synthesizing bash tools does not achieve full self-evolution, with multi-turn convergence results falling between the vanilla agent and \toolname{}, combining fine-grained bash commands into fixed-function scripts effectively reduces the length of the agent's repair trajectories and accelerates convergence.

Driven by these observations, we conduct further exploratory experiments by integrating Live-SWE-Agent with \toolname{}, aiming for a breakthrough in both performance and efficiency.

Specifically, we incorporate the prompt (as shown below) from Live-SWE-Agent, which instructs the model to synthesize new tools during the repair process, into the \toolname{} prompt while maintaining all other configurations. We evaluate this hybrid approach across three models (Qwen3.5-Plus, Devstral-24B, and DeepSeek-v3.1) on PATCHEVAL.
The experimental results indicate the following:

(1) Performance ceiling. The integration of Live-SWE-Agent does not significantly raise the upper bound of repair capabilities. Performance remains on par with the original results for Devstral-24B and DeepSeek-v3.1, while Qwen3.5-Plus successfully fixes three additional vulnerabilities.

(2) Model efficiency. The integration shows a measurable impact on efficiency. Both Qwen3.5-Plus and Devstral-24B reach convergence five turns earlier than their respective baselines.

Combined with the experimental results above, we believe that tool evolving and experience evolving are two critical pathways to achieving self-evolving agents. Among them, experience evolving focuses more on raising the performance ceiling, while tool evolving focuses more on enhancing efficiency. The organic integration of both can achieve a superior outcome that is both fast and effective.
\begin{tcolorbox}[
breakable,
colback=purple!5!white, 
colframe=purple!50!black,  
fonttitle=\bfseries, 
title=Tool evolving prompt in Live-SWE-Agent
]

\#\# Creating your own tools 

- You can also create your own tools in Python to help with your workflow

- Compared to basic bash commands, the tools you create should be able to better aid your workflow in solving the task

- Ensure each tool you create is in Python, contains informative outputs or error messages, and can be ran from the command line

- You should at least create a simple edit tool that can help you effectively edit arbitrary files instead of using bash commands

- The tools you create can be for any purpose, it does not need to be general, instead think about how it can help you specifically with the current task at hand

\#\#\# Example of creating a custom tool:

<example\_response>

THOUGHT: I noticed that in order to solve the issue I need to ... therefore I should create a custom tool to help me ...

```bash

cat <<'EOF' > /path/to/tool\_name.py

\#!/usr/bin/env python3

import sys

\# Import other packages if needed

def main():

    \quad\# Your tool logic here
    
    \quad...

if \_\_name\_\_ == "\_\_main\_\_":

    \quad main()
    
EOF

```

</example\_response>

\#\#\# Example of using the tool you created:

<example\_response>

THOUGHT: Let me use the custom tool I created to help me with ...

```bash

python /path/to/tool\_name.py <<EOF

your\_input\_here

EOF

```

\end{tcolorbox}
\section{Experience Transfer}\label{sec:exp_transfer_full}

As shown in Table~\ref{tab:transfer_result_full}, we present the full results of the experience transfer experiment. Specifically, we conduct experiments under three configurations, where the teacher model is consistently Qwen3.5-Plus, and the student models are Qwen3.5-Plus, Qwen3-Max, and Qwen3-Coder, respectively. We transfer the experience generated by \toolname{} using Qwen3.5-Plus on PATCHEVAL to VUL4J, and then compare it with the repair results without experience transfer under the same model. Overall, experience transfer demonstrates effectiveness across all models. Specifically, Qwen3.5-Plus fixes 3 additional vulnerabilities (31 $\to$ 34), Qwen3-Max fixes 2 additional vulnerabilities (23 $\to$ 25), and Qwen3-Coder-Plus fixes 1 additional vulnerability (22 $\to$ 23). By examining the repair details of each vulnerability in Table~\ref{tab:transfer_result_full}, we observe that the model’s repair behavior changes after applying experience transfer. Specifically, vulnerabilities that the model consistently fails to repair without prior knowledge can now be successfully fixed, as the introduced experience provides valuable guidance for the model to discover correct patches. Taking Qwen3.5-Plus as an example, the three vulnerabilities VUL4J-56, VUL4J-58, and VUL4J-71 cannot be fixed consistently by the model across multiple independent runs without experience guidance. However, after introducing experience transfer, the model achieves stable and reliable repair for these vulnerabilities. On the other hand, the introduced experience may also bring noise. For instance, Qwen3-Max can successfully repair VUL4J-25 without experience guidance but fails after applying experience transfer. This phenomenon is more pronounced in cross-model transfer (e.g., Qwen3.5-Plus $\to$ Qwen3-Max, Qwen3.5-Plus $\to$ Qwen3-Coder-Plus), while the effectiveness of experience is relatively robust under identical-model transfer (e.g., Qwen3.5-Plus $\to$ Qwen3-Plus). Besides the model choice, experience transfer is also affected by many other factors. For instance, the vulnerabilities in PATCHEVAL are written in JavaScript, Go, and Python, whereas VUL4J contains Java programs. The differences among programming languages impose a non‑negligible impact on the effectiveness of experience transfer. As a result, the performance of experience transfer across datasets, languages, and models degrades compared with experience evolution under the single dataset, same model, and identical language setting. In future work, we will further investigate how to improve the effectiveness of experience transfer. We plan to enhance its performance by extracting and transforming knowledge across programming languages, as well as by reducing the amount of experience to alleviate noise interference.

\input{tab/transfer_result_full}

\section{Case Study}

To better understand the efficacy of our experience-based approach, we conduct a systematic case study on vulnerabilities that \toolname{} successfully remediates but which remain intractable for vanilla agents. By examining diverse repair trajectories, ranging from successful strategies to lessons learned from past failures, we isolate the critical mechanisms through which our experience library enhances agent robustness and problem-solving efficiency.

\toolname{} leverages past failures as valuable lessons to avoid repeating mistakes in current repair trajectories. Taking the remediation of CVE-2020-8132 by GPT-5-mini as an example, by internalizing the failure experiences gained from the earlier repair of CVE-2020-7795, the model successfully resolves the vulnerability. In contrast, without such experience-based guidance, GPT-5-mini fails to remediate CVE-2020-8132. As illustrated in the figure below, we list the specific experiences utilized by \toolname{} when remediating CVE-2020-8132.

Furthermore, we observe how \toolname{} applies these positive experiences to novel challenges, such as the remediation of CVE-2022-0691 in the \texttt{url-parse} package. While the vanilla agent misdiagnoses the flaw as prototype pollution and attempts to over-engineer a complex wrapper, \toolname{} identifies the root cause (an incomplete whitespace regex) by referencing prior experience in robust input normalization. Guided by the experience rule, \textit{"Avoid regex/text manipulation for structured code edits; prefer AST-based or line-precise tools,"} \toolname{} replaces fragile \texttt{sed} scripts with precise Python-based line modification. This transition from brittle text processing to structurally sound edits ensures that the fix passes all test cases without inducing syntax errors.

The success of \toolname{} in these cases is attributed to three critical cognitive mechanisms:

\textbf{a. Shift from Blind Mutation to Structured Diagnostic Reasoning.}
While the vanilla agent reacts to patch application failures by blindly attempting complex, fragile regex transformations, ultimately inducing syntax errors, \toolname{} adopts a disciplined "suspend-diagnose-replan" workflow. By internalizing the rule to "test patch content before submission" and "verify workspace state," the agent effectively breaks the iterative error loop, transforming a failed deployment into a solvable debugging task.

\textbf{b. Enforcement of High-Confidence Security Paradigms.}
The library functions as a cognitive anchor against "repair depth degradation." When the vanilla agent faces prolonged difficulty, it regresses to superficial, unsafe fixes (e.g., simple regex-based whitelisting). In contrast, \toolname{} relies on explicit library guidance, such as \textit{"Prefer execFile over exec"}, to maintain a consistent security baseline, ensuring root-cause remediation of CWE-94.

\textbf{c. Mitigation of Environmental and Toolchain Entropy.}
A significant failure mode of the vanilla agent is "patch pollution," where irrelevant modifications (e.g., to \texttt{package-lock.json}) render patches non-applicable. \toolname{} mitigates this by strictly adhering to targeted patch-generation rules, such as \texttt{git diff -- . ':(exclude)*test*'}. By ensuring minimal, precise code edits, the agent preserves repository integrity, allowing automated verification to succeed. Both experiences used by \toolname{} are listed as follows.

\begin{tcolorbox}[
breakable,
colback=blue!5!white, 
colframe=blue!50!black,  
fonttitle=\bfseries, 
title=Failure experience from CVE-2020-7795 used by CVE-2020-8132
]

\textbf{Experience from CVE-2020-7795} \\
\textbf{CWE Types}: CWE-94, CWE-77, CWE-78 \\
\textbf{Quality Score}: 4/5 \\
\textbf{Similarity}: 0.84 \\
\textbf{Repair Result}: failure

\vspace{0.5em}
\textbf{Vulnerability Fix Report: CVE-2020-7795 (Command Injection in get-npm-package-version)}

\vspace{0.5em}
\textbf{A. Vulnerability Introduction and Analysis} \\
\textbf{Type}: Command Injection (CWE-77, CWE-78, CWE-94) \\
\textbf{Location}: \texttt{index.js}, lines 1–28 \\
\textbf{Root Cause}: The vulnerable code uses \texttt{child\_process.execSync()} with string interpolation to construct OS commands.
This allows an attacker to inject arbitrary shell commands via malicious \texttt{packageName} or \texttt{registry} values (e.g., \texttt{pkg; rm -rf /}). Since \texttt{execSync} passes the command string to a shell, special characters like \texttt{;}, \texttt{\&}, \texttt{|}, or \texttt{\$()} are interpreted, leading to arbitrary code execution. \\
\textbf{Impact}: Remote code execution in any application that uses this package with untrusted input.

\vspace{0.5em}
\textbf{B. Repair Strategy} \\
The correct mitigation strategy is to \textbf{avoid shell interpretation entirely} by:
\begin{enumerate}
    \item Using \texttt{child\_process.execFileSync()} instead of \texttt{execSync()}.
    \item Passing command arguments as an array (e.g., \texttt{['npm', 'view', packageName, 'version', ...]}) so inputs are treated as literal arguments, not shell tokens.
    \item Adding basic input validation to reject obviously malicious patterns (defense-in-depth).
\end{enumerate}
This approach eliminates the injection surface because the OS executes \texttt{npm} directly with arguments, bypassing the shell.

\vspace{0.5em}
\textbf{C. Trajectory Analysis} \\
\textbf{Positive Analysis}:
\begin{itemize}
    \item \textbf{Correct identification of root cause}: The assistant correctly recognized that \texttt{execSync} with string interpolation was the vulnerability source.
    \item \textbf{Adoption of safe alternative}: Switching to \texttt{execFileSync} with argument arrays is the standard, secure pattern for spawning subprocesses in Node.js.
    \item \textbf{Input validation added}: While not strictly necessary if shell is avoided, validating \texttt{packageName} and \texttt{registry} against allowlists (e.g., $/\text{\textasciicircum}[\text{\textbackslash}w@\text{\textbackslash}/\text{\textbackslash}-.]+[\$]/$) provides defense-in-depth.
\end{itemize}
\textbf{Negative Analysis}:
\begin{itemize}
    \item \textbf{Patch generation errors}: The assistant repeatedly included unintended changes (e.g., to \texttt{package-lock.json}) in \texttt{fix.patch}, causing \texttt{fix-run.sh} to fail during patch application. This indicates a lack of precision in generating diffs.
    \item \textbf{Incorrect file path handling}: In some attempts, the assistant wrote to \texttt{index.js} in the wrong directory or failed to isolate changes to only the vulnerable file.
    \item \textbf{Overcomplicated shell commands}: Using complex here-docs with quoting issues led to syntax errors (Turn 6), delaying progress.
\end{itemize}

\vspace{0.5em}
\textbf{D. Experience Summary} \\
\textbf{Rule 1: Prefer \texttt{execFile}/\texttt{spawn} over \texttt{exec} for external commands} \\
Applicability: Node.js, CWE-77/78; Goal: Prevent shell injection; Guidance: Never use \texttt{exec()} or \texttt{execSync()} with user input. Always use \texttt{execFile()} or \texttt{spawn()} with argument arrays. 
UNSAFE: $\text{execSync}(`\text{npm view } \$\{pkg\} \text{ version}`);$
SAFE: $\text{execFileSync}(\text{'npm'}, [\text{'view'}, pkg, \text{'version'}]);$ \\

\textbf{Rule 2: Generate minimal, targeted patches} \\
Applicability: General, all languages; Goal: Ensure patch applies cleanly during validation; Guidance: Use \texttt{git diff -- . ':(exclude)*test*'} to exclude unrelated files. Verify \texttt{fix.patch} contains only intended changes before submission. \\

\textbf{Rule 3: Validate input even when using safe APIs} \\
Applicability: Defense-in-depth, Node.js; Goal: Reduce attack surface; Guidance: Apply allowlist regexes to inputs like package names (e.g., $/\text{\textasciicircum}[\text{\textbackslash}w@\text{\textbackslash}/\text{\textbackslash}-.]+[\$]/$). Reject empty or non-string inputs early.

\vspace{0.5em}

\textbf{E. Reflection and Improvement} \\
Although the assistant correctly identified the fix strategy (use \texttt{execFileSync} + validation), the repair \textbf{failed} due to operational errors in patch generation and environment handling. Future efforts should:
\begin{itemize}
    \item Test patch content before submission (\texttt{cat fix.patch} to inspect).
    \item Ensure only the target file is modified.
    \item Avoid modifying \texttt{package-lock.json} unless absolutely necessary.
    \item Use \texttt{git status} and \texttt{git diff} to verify workspace state.
\end{itemize}
Peer review of patches and incremental testing would improve success rates. The core remediation logic was sound—only the delivery mechanism failed.

\end{tcolorbox}

\begin{tcolorbox}[
breakable,
colback=blue!5!white, 
colframe=blue!50!black,  
fonttitle=\bfseries, 
title=Successful experience from CVE-2022-0512 used by CVE-2022-0691
]

\textbf{Experience from CVE-2022-0512} \\
\textbf{CWE Types}: CWE-862, CWE-639 \\
\textbf{Quality Score}: 4/5 \\
\textbf{Similarity}: 1.00 \\
\textbf{Repair Result}: success

\vspace{0.5em}
\textbf{Vulnerability Fix Report: CVE-2022-0512 in url-parse}

\vspace{0.5em}
\textbf{A. Vulnerability Introduction and Analysis} \\
\textbf{CVE ID}: CVE-2022-0512 \\
\textbf{CWE Types}: CWE-862 (Missing Authorization), CWE-639 (Authorization Bypass Through User-Controlled Key) \\
\textbf{Language}: JavaScript (Node.js / NPM package \texttt{url-parse}) \\
\textbf{Location}: \texttt{url-parse/index.js}, lines 236–389 and 402–488 \\
\textbf{Description}: The vulnerability allows an attacker to bypass authorization by manipulating the \texttt{auth} component of a URL (e.g., \texttt{user@:pass@}). When parsed, special characters like \texttt{@} or \texttt{:} in username/password fields were not properly percent-encoded, leading to incorrect parsing. \\
\textbf{Root Cause}: In both the constructor (\texttt{Url}) and the \texttt{set('auth', value)} method:
\begin{itemize}
    \item The \texttt{auth} string was split naively using \texttt{.split(':')}, which fails when passwords contain colons.
    \item Username and password values were assigned \textbf{without percent-encoding}, allowing raw \texttt{@} or \texttt{:} to interfere with URL structure.
    \item No normalization was applied to handle already-encoded inputs, risking double-encoding or inconsistent behavior.
\end{itemize}
\textbf{Impact}: An attacker could craft a URL like \texttt{http://attacker@:victim-password/} that, when parsed, incorrectly assigns \texttt{attacker@} as the username and \texttt{victim-password} as the password.

\vspace{0.5em}
\textbf{B. Repair Strategy} \\
The fix ensures that:
\begin{enumerate}
    \item \textbf{Auth is split only on the first colon}: Correctly separates username from password.
    \item \textbf{Username and password are safely normalized}: Existing percent-encodings are decoded first (with error tolerance), then re-encoded using \texttt{encodeURIComponent}.
\end{enumerate}
This strategy preserves backward compatibility while eliminating the injection surface, addressing CWE-639 by ensuring user-controlled keys cannot inject structural characters.

\vspace{0.5em}
\textbf{C. Trajectory Analysis} \\
\textbf{Positive Analysis}:
\begin{itemize}
    \item \textbf{Test-driven validation}: The assistant repeatedly ran \texttt{fix-run.sh} to ensure behavioral correctness.
    \item \textbf{Safe encoding logic}: Introduced a \texttt{safeDecode} helper with \texttt{try/catch} to avoid crashes.
    \item \textbf{First-colon split}: Replaced \texttt{.split(':')} with \texttt{indexOf(':')} + \texttt{slice()}.
\end{itemize}
\textbf{Negative Analysis}:
\begin{itemize}
    \item \textbf{Initial over-encoding}: Early attempts applied \texttt{encodeURIComponent} without decoding first, causing double-encoding.
    \item \textbf{Fragile patching}: Used complex \texttt{perl -pe} and regex-based replacements that broke due to quoting issues.
    \item \textbf{Assumed file paths}: Repeatedly tried to access files from incorrect directories.
\end{itemize}

\vspace{0.5em}
\textbf{D. Experience Summary} \\
\textbf{Rule 1: Normalize User Input via Decode-Then-Encode} \\
Guidance: When accepting possibly pre-encoded strings, \textbf{always decode first (safely), then re-encode}.
\[
\begin{aligned}
\text{safeNormalize}(s) = \text{try } &\{ \text{encodeURIComponent} \\
&(\text{decodeURIComponent}(s)) \} \\
& \text{catch } \{ \text{encodeURIComponent}(s) \}
\end{aligned}
\]

\textbf{Rule 2: Split on First Occurrence for Delimited Fields} \\
Guidance: Use \texttt{indexOf(delimiter)} + \texttt{slice()}, not \texttt{split(delimiter)}.

\smallskip
\noindent\texttt{idx = str.indexOf(':');}\\
\texttt{user = idx == -1 ? str : str.slice(0, idx);}

\textbf{Rule 3: Validate Against Real Test Cases Early} \\
Guidance: Locate and inspect the \textbf{exact failing test assertions} before implementing a fix.

\vspace{0.5em}

\textbf{E. Reflection and Improvement} \\
Future efforts should avoid regex/text manipulation for code edits and prefer more precise tools. The final approach achieves idempotency ($f(f(x)) \equiv f(x)$), which is essential for robust normalization.

\end{tcolorbox}

\section{Prompt Design}

we detail the prompts utilized in our experiments for both the vanilla agent and \toolname{} as follows. The system prompt is shared across all tasks, while the instance prompt is dynamically constructed for each remediation session based on specific vulnerability details. Additionally, the experience prompt is unique to \toolname{}.

\begin{tcolorbox}[
  breakable, 
  colback=green!5!white, 
  colframe=green!50!black, 
  fonttitle=\bfseries, 
  title=Prompts used by \toolname{}
]

\textbf{System Prompt}

You are a helpful assistant that is equipped with tools to solve programming tasks.

Your response must contain exactly ONE tool call.

Include a THOUGHT section in your response where you explain your reasoning process.\\

\textbf{Instance Prompt}

<cve\_description>

The vulnerability has the following CVE description:

\{cve\_description\}

</cve\_description>
\\

<cwe\_info>

The vulnerability has the following CWE information:

\{cwe\_info\}

</cwe\_info>
\\

\{\% if experience\_context \%\}

<historical\_experiences>

The following are relevant experiences from similar vulnerability fixes .

These experiences are extracted from successful past fixes. Use them as reference to understand common patterns,

but adapt the solutions to the current context rather than copying them directly:

\{experience\_context\}

</historical\_experiences>

\{\% endif \%\}
\\

<vul\_location>

The vulnerability is located in the following code lines, do not modify other code lines:

\{vul\_location\}

</vul\_location>\\

<instructions>\\

TASK INSTRUCTIONS OVERVIEW

You're a software engineer interacting continuously with a computer by calling tools.
You'll be helping implement necessary changes to fix the vulnerability.
Your task is specifically to make changes to non-test files in the current directory 
in order to fix the vulnerability described above in a way that is general and consistent with the codebase.
To enhance your vulnerability remediation capabilities and experience, you should maintain a dedicated documentation 
for each vulnerability. This documentation should record details such as the vulnerability description, 
root cause analysis, remediation process, and outcomes. You can refer to previously documented records 
during the remediation process and continuously update and refine them for future reference.\newline

IMPORTANT: This is an interactive process where you will think and call ONE tool, see its result, then think and call your next tool.\\

For each response:

1. Include a THOUGHT section explaining your reasoning and what you're trying to accomplish\\

2. Provide exactly ONE tool to call\\

IMPORTANT BOUNDARIES\\

- MODIFY: Regular source code files

- DO NOT MODIFY: Tests, configuration files (pyproject.toml, setup.cfg, etc.)\\

AVAILABLE SCRIPTS\\

The following four scripts are available to help you fix and verify the vulnerability:

bash fix-run.sh: Run fix-run.sh to execute the PoC and verify that the patched vulnerability is no longer exploitable.
                To run fix-run.sh, you must first generate fix.patch in the same directory.
                
bash vul-run.sh: Run vul-run.sh to execute the PoC and verify that the original (unpatched) vulnerability is exploitable.

bash unit\_test.sh: (if present) Run unit\_test.sh to execute unit tests and validate functional correctness.

bash prepare.sh: Resets all changes in the repository. Run this script before each evaluation.\\

RECOMMENDED WORKFLOW\\

1. Understand the vulnerability from the CVE description and CWE information

2. Check the latest documentation of the vulnerability if available

3. Verify the existence of the vulnerability by running vul-run.sh

4. Locate vulnerability in the source code if needed

5. Edit source code to fix the vulnerability

6. Use git diff to review your changes, generate fix.patch, and use cat to view its content

7. Verify fix.patch by running prepare.sh and fix-run.sh, if test error occurs, you can check test.patch for reference

8. If fix.patch passes verification, update documentation, submit fix.patch and finish workflow

Each response should include:

1. A THOUGHT section where you explain your reasoning and plan

2. A single tool call\\

CRITICAL REQUIREMENTS:\\

- Your response SHOULD include a THOUGHT section explaining your reasoning

- Your response MUST include EXACTLY ONE tool call\\

ENVIRONMENT DETAILS\\

- You have a full Linux shell environment

- Always use non-interactive flags (-y, -f) for commands

- Avoid interactive commands like vi, nano, or any that require user input

- If a command isn't available, you can install it

- Source code dir is \{source\_dir\}, it is a git repository; you can use git commands to inspect changes

- The test.patch file documents the new tests that should pass after the vulnerability is fixed.\\

USEFUL COMMAND EXAMPLES\\

\textbf{Create a new file}

\textit{cat << 'EOF' > newfile.py}

\textit{import numpy as np}

\textit{hello = "world"}

\textit{print(hello)}

\textit{EOF}\\

\textbf{Edit files with sed}

\textbf{a. Replace all occurrences}

\textit{sed -i 's/old\_string/new\_string/g' filename.py}

\textbf{b. Replace only first occurrence}

sed -i 's/old\_string/new\_string/' filename.py

\textbf{c. Replace first occurrence on line 1}

sed -i '1s/old\_string/new\_string/' filename.py

\textbf{d. Replace all occurrences in lines 1-10}

sed -i '1,10s/old\_string/new\_string/g' filename.py\\

\textbf{View file content}

nl -ba filename.py | sed -n '10,20p'\\

\textbf{Generate diff of changes}

cd \{source\_dir\} \&\& git diff -- . ':(exclude)*test*' ':(exclude)*Test*' > ../fix.patch\\

\textbf{Validate the patch}

prepare.sh is used to reset all changes, run it before validation.

please do not run unit tests for verification by yourself. Use fix-run.sh instead, because test.patch has not been included.

\textit{bash prepare.sh \&\& bash fix-run.sh}\\

Any other command you want to run

\textit{anything}

</instructions>
\end{tcolorbox}

\section{Early Stop}
To balance the performance and cost of \toolname{}, we have integrated an early-stopping strategy into our original methodology. The core rationale behind this strategy is that when the remediation cost of a single turn for \toolname{} outweighs the marginal gain from new fixes beyond a certain threshold, we deem the repair process no longer cost-effective and thus terminate the entire workflow prematurely. Specifically, we define the turn-level yield rate $\alpha$ (detailed in Equation~\ref{eq:alpha}) as the ratio between the marginal increase in repair rate and the marginal increase in cost. When $\alpha$ falls below a certain threshold $t$, \toolname{} triggers an early-stopping strategy. To facilitate the determination of the optimal threshold for $\alpha$, Table~\ref{tab:early_stop_detail} provides a detailed breakdown of \toolname{}’s turn-level computational costs and incremental fixes, based on which the corresponding values of $\alpha$ are calculated. As shown, $\alpha$ exhibits a generally downward trend as the number of turns increases. By adjusting the threshold $t$, \toolname{} can achieve an optimal balance between performance and cost-effectiveness across different models. Based on the empirical data in the table, we recommend setting the threshold $t$ within the range of 0.15 to 0.25.

\input{tab/early_stop_detail}

\end{document}

%% file: tab/overall_performance.tex

\begin{table*}[htbp]
\centering
\caption{Performance comparison of different vulnerability repair methods on PATCHEVAL and SEC-bench. VulRepair and VulMaster use CodeT5; OpenHands and Claude Code use GPT-5; other baselines use GPT-5-mini. Best results are in bold.}
\label{tab:vul_repair_main}
\small
\setlength{\tabcolsep}{4.5pt}
\renewcommand{\arraystretch}{1.2}

\resizebox{1.0\linewidth}{!}{\begin{tabular}{llcccccccccccccccc}
\toprule
\multirow{2}{*}{Dataset} & \multirow{2}{*}{Metric}
& \multicolumn{2}{c}{Learning-based}
& \multicolumn{6}{c}{LLM-based}
& \multicolumn{7}{c}{Agent-based}
& Ours \\
\cmidrule(lr){3-4}
\cmidrule(lr){5-10}
\cmidrule(lr){11-17}
\cmidrule(l){18-18}
&
& \makecell[c]{VulRepair}
& \makecell[c]{VulMaster}
& \makecell[c]{Few-Shot}
& \makecell[c]{Zero-Shot}
& \makecell[c]{IntentFix}
& \makecell[c]{PailGen}
& \makecell[c]{ChatRepair}
& \makecell[c]{LoopRepair}
& \makecell[c]{SmolAgent}
& \makecell[c]{AgentMem}
& \makecell[c]{OpenHands}
& \makecell[c]{Claude\\Code}
& \makecell[c]{Vanilla}
& \makecell[c]{Live-SWE}
& \makecell[c]{ContraFix}
& \textbf{\toolname{}} \\
&& FSE 2022 & ICSE 2024 & 2020 & 2022 & ICSE 2026 & TOSEM 2026 & ISSTA 2024 & ICSE 2026 & 2025 & 2025 & 2024 & 2025 & 2026 & 2026 & 2026 & 2026 \\
\midrule
\multirow{2}{*}{\makecell[c]{PATCHEVAL\\(230)}}
& \# Fix     & 0 & 0 & 39 & 34 & 52 & 52 & 115 & 124 & - & - & 84 & 80 & 194 & 199 & 166 & \textbf{215}\\
& \% Fix     & 0.00 & 0.00 & 16.95 & 14.78 & 22.61 & 22.61 & 50.00 & 53.91 & - & - & 36.52 & 34.78 & 84.34 & 86.52 & 72.17 & \textbf{93.47}\\
\addlinespace[2pt]
\multirow{2}{*}{\makecell[c]{SEC-bench\\(200)}}
& \# Fix     & 7 & 19 & 60 & 69 & 73 & 81 & 103 & 107 & 69 & 100 & - & - & 155 & 160 & 168 & \textbf{174}\\
& \% Fix     & 3.50 & 9.50 & 30.00 & 34.50 & 36.50 & 40.50 & 51.50 & 53.50 & 34.50 & 50.00 & - & - & 77.50 & 80.00 & 84.00 & \textbf{87.00}\\
\midrule
\multirow{2}{*}{\makecell[c]{Total\\(430)}}
& \# Fix & 7 & 19 & 99 & 103 & 125 & 133 & 218 & 231 & - & - & - & - & 349 & 359 & 334 & \textbf{389}\\
& \% Fix & 1.62 & 4.41 & 23.02 & 23.95 & 29.06 & 30.93 & 50.69 & 53.72 & - & - & - & - & 81.16 & 83.48 & 77.67 & \textbf{90.46}\\
\bottomrule
\end{tabular}}
\label{tab:overall_performance}
\end{table*}

%% file: tab/ablation_score.tex
\begin{table}[htbp]
  \centering
  \caption{Effectiveness of the experience retrieval mechanism (w/ experience score vs. w/o experience score).}
    \resizebox{0.6\linewidth}{!}{\begin{tabular}{lrrr}
    \toprule
    Model  & w/o score & w/ score \\
    \midrule
    GPT-5-mini    & 205   & 215 (\textcolor{deepgreen}{+10}) \\
    DeepSeek-v3.1    & 160   & 163 (\textcolor{deepgreen}{+3}) \\
    Qwen3.5-Plus    & 208   & 220 (\textcolor{deepgreen}{+12}) \\
    Devstral-24B    & 213   & 218 (\textcolor{deepgreen}{+5}) \\
    Devstral-123B    & 215   & 216 (\textcolor{deepgreen}{+1}) \\
    \bottomrule
    \end{tabular}}%
  \label{tab:ablation_score}%
\end{table}%

%% file: tab/ablation_cold_start.tex
\begin{table}[htbp]
  \centering
  \caption{First-turn repair results under different cold-start strategies.}
    \resizebox{0.75\linewidth}{!}{\begin{tabular}{lrrr}
    \toprule
    Model & w/o cold-start & Standard Patch & Pre-repair \\
    \midrule
    GPT-5-mini & 78    & 57 (\textcolor{deepred}{-21}) & 105 (\textcolor{deepgreen}{+27}) \\
    DeepSeek-v3.1 & 53    & 46 (\textcolor{deepred}{-7}) & 54 (\textcolor{deepgreen}{+1}) \\
    Qwen3.5-Plus & 100   & 110 (\textcolor{deepgreen}{+10}) & 131 (\textcolor{deepgreen}{+31}) \\
    Devstral-24B & 111   & 120 (\textcolor{deepgreen}{+9}) & 113 (\textcolor{deepgreen}{+2}) \\
    Devstral-123B & 119   & 128 (\textcolor{deepgreen}{+9}) & 124 (\textcolor{deepgreen}{+5}) \\
    \bottomrule
    \end{tabular}}%
  \label{tab:ablation_cold_start}%
\end{table}%

%% file: tab/vul4j.tex
\begin{table*}[hthp]
\centering
\footnotesize
\caption{The effectiveness of experience transfer (PATCHEVAL $\to$ VUL4J).}
\resizebox{0.8\linewidth}{!}{
\begin{tabular}{lrrrrrrr}
\toprule
Method        & \toolname{}-T1 & \toolname{}-T2 & NTR~\cite{ntr} & VRPILOT~\cite{vrpilot} & APR4Vul~\cite{bui2024apr4vul} & ChatRepair~\cite{xia2024automated} & TSAPR~\cite{hu2025tsapr}\\ 
\midrule
\# Fix  &  34/79 (\textcolor{deepgreen}{+9.67\%})    &   25/79 (\textcolor{deepgreen}{+8.69\%}) & 14/79  & 14/79 & 16/79 & 14/79 & 24/79 \\
\bottomrule 
\end{tabular}}
\label{tab:vul}
\end{table*}

%% file: tab/cost.tex
\begin{table*}[htbp]
  \centering
  \caption{Cost analysis of \toolname{} on PATCHEVAL.}
    \resizebox{0.9\linewidth}{!}{\begin{tabular}{llrrrrrrrr}
    \toprule
    \multirow{2}[4]{*}{Model} & \multirow{2}[4]{*}{Method} & \multicolumn{4}{r}{Per CVE}   & \multicolumn{3}{r}{Total} & Ref \\
\cmidrule(lr){3-6}  \cmidrule(lr){7-9} \cmidrule(lr){10-10}        &       & Step  & Prompt (K) & Response (K) & Turn  & Prompt (M) & Response (M) & Cost (\$) & Charge (\$/M) \\
    \midrule
    \multirow{2}[1]{*}{DeepSeekv3.1} & Vanilla & 64.15 & 823.8 & 11.4  & 3.51  & 147.7  & 20.4  & 419.2 & Prompt: 0.27 \\
          & EvoRepair &  76.18     &  1355.7     &   10.6    &  3.93     &  214     &  16.7     &  594.7     & Response: 1 \\
    \multirow{2}[0]{*}{GPT-5-mini} & Vanilla & 20.02 & 306.4 & 33.3  & 3.03  & 337.9 & 36.7  & 173.8 & Prompt: 0.275 \\
          & EvoRepair & 25.64 & 596.3 & 43.9  & 2.49  & 444.8 & 32.7  & 194.4 & Response: 2.2 \\
    \multirow{2}[0]{*}{Devstral-24B} & Vanilla & 68.99 & 1321.7 & 10.3  & 2.33  & 925.2 & 7.1   & 94.6 & Prompt: 0.1 \\
          & EvoRepair & 66.22 & 1587.1 & 11.1  & 2.41  & 1101.4 & 7.7   & 112.5 & Response: 0.3 \\
    \multirow{2}[0]{*}{Devstral-123B} & Vanilla & 73.04 & 1203  & 9.5   & 1.99  & 715.8 & 5.6   & 297.6 & Prompt: 0.4 \\
          & EvoRepair & 65.25 & 1269.7 & 7.4   & 2.31  & 883.6 & 5.1   & 363.8 & Response: 2 \\
    \multirow{2}[1]{*}{Qwen3.5-Plus} & Vanilla & 82.01 & 1881.6 & 20.6  & 2.56  & 1580.5 & 20.6  & 681.79 & Prompt: 0.4 \\
          & EvoRepair & 70.28 & 1681.6 & 26    & 1.92  & 911.4 & 14    & 398.3 & Response: 2 \\
    \bottomrule
    \end{tabular}}%
  \label{tab:cost_analysis}%
\end{table*}%

%% file: tab/early_stop_summary.tex
\begin{table}[htbp]
  \centering
  \caption{Early stop (ES) strategy under different $\alpha$.}
    \resizebox{1.0\linewidth}{!}{\begin{tabular}{llrrrrrr}
    \toprule
    \multicolumn{1}{l}{Model} & $\alpha$ & 0 (w/o ES)     & 0.1   & 0.15  & 0.2   & 0.25  & 0.3 \\
    \midrule
    \multicolumn{1}{l}{\multirow{2}[1]{*}{DeepSeek-v3.1}} & \# Fix & 163   & 163(-) & 163(-) & 150(\textcolor{deepred}{-7.97\%}) & 150(\textcolor{deepred}{-7.97\%}) & 91(\textcolor{deepred}{-44.17\%}) \\
          & Cost (\$) & 594.74 & 594.74(-) & 594.74(-) & 474.63(\textcolor{deepred}{-20.19\%}) & 474.63(\textcolor{deepred}{-20.19\%}) & 200.18(\textcolor{deepred}{-66.34\%}) \\
    \multicolumn{1}{l}{\multirow{2}[0]{*}{Qwen3.5-Plus}} & \# Fix & 220   & 220(-) & 207(\textcolor{deepred}{-5.9\%}) & 207(\textcolor{deepred}{-5.9\%}) & 207(\textcolor{deepred}{-5.9\%}) & 205(\textcolor{deepred}{-6.8\%}) \\
          & Cost (\$) & 398.35 & 398.35(-) & 294.92(\textcolor{deepred}{-25.96\%}) & 294.92(\textcolor{deepred}{-25.96\%}) & 294.92(\textcolor{deepred}{-25.96\%}) & 274.43(\textcolor{deepred}{-31.1\%}) \\
    \multicolumn{1}{l}{\multirow{2}[0]{*}{Devstral-123B}} & \# Fix & 216   & 208(\textcolor{deepred}{-3.7\%}) & 195(\textcolor{deepred}{-9.72\%}) & 195(\textcolor{deepred}{-9.72\%}) & 195(\textcolor{deepred}{-9.72\%}) & 195(\textcolor{deepred}{-9.72\%}) \\
          & Cost (\$) & 363.81 & 314.95(\textcolor{deepred}{-13.43\%}) & 246.61(\textcolor{deepred}{-32.21\%}) & 246.61(\textcolor{deepred}{-32.21\%}) & 246.61(\textcolor{deepred}{-32.21\%}) & 246.61(\textcolor{deepred}{-32.21\%}) \\
    \multicolumn{1}{l}{\multirow{2}[0]{*}{Devstral-24B}} & \# Fix & 218   & 218(-) & 214(\textcolor{deepred}{-1.83\%}) & 195(\textcolor{deepred}{-10.5\%}) & 195(\textcolor{deepred}{-10.55\%}) & 195(\textcolor{deepred}{-10.55\%}) \\
          & Cost (\$) & 112.45 & 112.45(-) & 102.21(\textcolor{deepred}{-9.1\%}) & 75.86(\textcolor{deepred}{-32.53\%}) & 75.86(\textcolor{deepred}{-32.53\%}) & 75.86(\textcolor{deepred}{-32.53\%}) \\
    \multicolumn{1}{l}{\multirow{2}[1]{*}{GPT-5-mini}} & \# Fix & 215   & 204(\textcolor{deepred}{-5.11\%}) & 204(\textcolor{deepred}{-5.1\%}1) & 202(\textcolor{deepred}{-6.04\%}) & 198(\textcolor{deepred}{-7.9\%}) & 198(\textcolor{deepred}{-7.9\%}) \\
          & Cost (\$) & 194.4 & 147.53(\textcolor{deepred}{-24.11\%}) & 147.53(\textcolor{deepred}{-24.11\%}) & 132.92(\textcolor{deepred}{-31.62\%}) & 120.49(\textcolor{deepred}{-38.01\%}) & 120.49(\textcolor{deepred}{-38.01\%}) \\
    \bottomrule
    \end{tabular}}%
  \label{tab:early_stop_summary}%
\end{table}%

%% file: tab/algorithm.tex
\begin{algorithm}[htbp]
\caption{Self-Evolving Framework for Automated Vulnerability Repair}
\label{alg:self_evolving_framework}
\KwIn{Current vulnerability \( q \), initial experience bank \( E \)}
\KwOut{Final patched code}

\Begin{
    // Initialize variables\\
    \( T \gets 0 \) \tcp*{Repair turns}
    \( V \gets \text{Unfixed vulnerabilities} \)\\

    \While{\( V \) is not empty}{
        \( T \gets T + 1 \)\\
        
        // Step 1: Retrieve relevant experiences\\
        \( S \gets \text{Experience Retrieval}(q, E) \)\\
        
        // Step 2: Prioritize retrieval based on quality and similarity\\
        \( R \gets \text{Prioritize}(S) \)\\
        
        // Step 3: Inject retrieved experiences into the repair context\\
        \( \text{Inject}(R) \)\\

        // Step 4: Perform repair using the vanilla agent\\
        \( \text{Patch} \gets \text{VanillaAgentRepair}(q) \)\\

        // Step 5: Validate patch and update experience bank\\
        \If{\text{Validation}(Patch, q)}{
            \( E \gets \text{UpdateExperience}(Patch, T) \)\\
            \( V \gets V \backslash \{q\} \) \tcp*{Remove fixed vulnerability}
        }
    }
    
    \Return \text{Final patched code} 
}
\end{algorithm}

%% file: tab/secbench_tts.tex
\begin{table*}[htbp]
  \centering
  \caption{Comparison of test-time scaling performance (\toolname{}, Live-SWE-Agent, Vanilla Agent) on SEC-bench.}
    \resizebox{1.0\linewidth}{!}{\begin{tabular}{lccccccccccccccccc}
    \toprule
    \rowcolor[rgb]{ .906,  .902,  .902} Model & Method & Turn 1 & Turn 2 & Turn 3 & Turn 4 & Turn 5 & Turn 6 & Turn 7 & Turn 8 & Turn 9 & Turn 10 & Turn 11 & Turn 12 & Turn 13 & Turn 14 & Turn 15 & \# Fix \\
    \midrule

    \rowcolor[rgb]{ .886,  .937,  .855}       & Vanilla & 22    & 50    & 65    & 75    & 83    & 90    & 95    & 98    & 99    & 101   & 111   & 114   & 117   & 118   & 118   & 118 \\
    \rowcolor[rgb]{ .886,  .937,  .855} DeepSeek-v3.1 & Live-SWE-Agent & 28    & 59    & 73    & 83    & 95    & 102   & 108   & 112   & 120   & 121   & 125   & 127   & 127   & - & - & 127 \\
    \rowcolor[rgb]{ .886,  .937,  .855}       & \toolname{} & 44    & 73    & 90    & 102   & 110   & 112   & 116   & 120   & 125   & 126   & 131   & 131   & - & - & - & 131 \\

    \rowcolor[rgb]{ .776,  .878,  .706}       & Vanilla & 94    & 122   & 140   & 149   & 149   & - & - & - & - & - & - & - & - & - & - & 149 \\
    \rowcolor[rgb]{ .776,  .878,  .706} DeepSeek-v3.2 & Live-SWE-Agent & 86    & 122   & 135   & 146   & 150   & 154   & 159   & 163   & 166   & 167   & 171   & 173   & 173   & - & - & 173 \\
    \rowcolor[rgb]{ .776,  .878,  .706} & \toolname{} & 101   & 134   & 147   & 156   & 163   & 168   & 169   & 171   & 173   & 173   & - & - & - & - & - & 173 \\

    \rowcolor[rgb]{ 1,  .949,  .8}       & Vanilla & 36    & 60    & 71    & 77    & 78    & 82    & 87    & 89    & 94    & 94    & - & - & - & - & - & 94 \\
    \rowcolor[rgb]{ 1,  .949,  .8} Qwen3.5-Plus & Live-SWE-Agent & 67    & 84    & 101   & 111   & 116   & 121   & 125   & 126   & 128   & 131   & 134   & 137   & 138   & 141   & 144   & 144 \\
    \rowcolor[rgb]{ 1,  .949,  .8}       & \toolname{} & 64    & 98    & 120   & 135   & 141   & 146   & 149   & 152   & 156   & 159   & 165   & 168   & 170   & 170   & - & 170 \\
    
    \rowcolor[rgb]{ .867,  .922,  .969}       & Vanilla & 51    & 81    & 98    & 110   & 122   & 128   & 131   & 138   & 142   & 145   & 150   & 152   & 153   & 155   &   155    & 155 \\
    \rowcolor[rgb]{ .867,  .922,  .969} GPT-5-mini & Live-SWE-Agent & 35    & 74    & 102   & 117   & 125   & 132   & 141   & 145   & 148   & 151   & 152   & 156   & 157   & 160   & 160   & 160 \\
    \rowcolor[rgb]{ .867,  .922,  .969}       & \toolname{} & 66    & 119   & 135   & 142   & 149   & 151   & 155   & 158   & 161   & 164   & 166 & 170 & 171 & 174 & 174 & 174 \\

    \bottomrule
    \end{tabular}}%
  \label{tab:tts_sec}%
\end{table*}%

%% file: tab/secbench_file_loc.tex
\begin{table*}[htbp]
  \centering
  \caption{File-level localization accuracy on SEC-bench.}
    \resizebox{1.0\linewidth}{!}{\begin{tabular}{lcccccccccccccccc}
    \toprule
    \rowcolor[rgb]{ .906,  .902,  .902} Model & Method & Turn 1 & Turn 2 & Turn 3 & Turn 4 & Turn 5 & Turn 6 & Turn 7 & Turn 8 & Turn 9 & Turn 10 & Turn 11 & Turn 12 & Turn 13 & Turn 14 & Turn 15 \\
    \midrule

    \rowcolor[rgb]{ .886,  .937,  .855}  & Vanilla & 49.07\% & 41.76\% & 43.55\% & 38.60\% & 41.46\% & 40.82\% & 45.45\% & 35.42\% & 30.43\% & 51.52\% & 52.17\% & 25.45\% & 54.17\% & 50.00\% & 32.14\% \\
    \rowcolor[rgb]{ .886,  .937,  .855}    DeepSeek-v3.1   & Live-SWE-Agent & 41.28\% & 39.60\% & 33.77\% & 42.86\% & 27.50\% & 39.47\% & 48.15\% & 25.93\% & 33.96\% & 44.00\% & 35.71\% & 25.00\% & 35.14\% & -     & - \\
    \rowcolor[rgb]{ .886,  .937,  .855}       & \toolname{} & 49.69\% & 47.75\% & 41.84\% & 45.45\% & 37.65\% & 39.39\% & 41.43\% & 48.28\% & 42.86\% & 36.54\% & 40.35\% & 36.67\% & -     & -     & - \\

    \rowcolor[rgb]{ .776,  .878,  .706}  & Vanilla & 46.67\% & 52.88\% & 56.06\% & 48.00\% & 46.81\% & -     & -     & -     & -     & -     & -     & -     & -     & -     & - \\
    \rowcolor[rgb]{ .776,  .878,  .706}   DeepSeek-v3.2    & Live-SWE-Agent & 47.08\% & 52.73\% & 48.15\% & 52.24\% & 50.00\% & 47.37\% & 50.00\% & 41.18\% & 59.26\% & 41.18\% & 44.83\% & 57.69\% & 52.63\% & -     & - \\
    \rowcolor[rgb]{ .776,  .878,  .706}       & \toolname{} & 47.26\% & 45.76\% & 35.29\% & 38.71\% & 40.74\% & 42.50\% & 63.33\% & 46.43\% & 50.00\% & 47.83\% & -     & -     & -     & -     & - \\

    \rowcolor[rgb]{ 1,  .949,  .8}  & Vanilla & 41.18\% & 61.90\% & 42.42\% & 41.38\% & 38.46\% & 47.37\% & 57.89\% & 28.57\% & 54.17\% & 63.64\% & -     & -     & -     & -     & - \\
    \rowcolor[rgb]{ 1,  .949,  .8}   Qwen3.5-Plus    & Live-SWE-Agent & 62.16\% & 38.33\% & 53.33\% & 57.14\% & 57.89\% & 47.37\% & 66.67\% & 46.15\% & 62.50\% & 50.00\% & 63.64\% & 53.85\% & 54.55\% & 87.50\% & 60.00\% \\
    \rowcolor[rgb]{ 1,  .949,  .8}       & \toolname{} & 53.23\% & 54.79\% & 55.81\% & 50.00\% & 65.22\% & 61.11\% & 27.78\% & 33.33\% & 72.73\% & 53.33\% & 53.33\% & 37.50\% & 60.00\% & 40.00\% & - \\

    \rowcolor[rgb]{ .867,  .922,  .969}  & Vanilla & 30.94\% & 32.37\% & 26.90\% & 25.62\% & 27.64\% & 23.64\% & 25.00\% & 30.85\% & 22.58\% & 20.43\% & 25.88\% & 22.64\% & 25.42\% & 36.73\% & 21.57\% \\
    \rowcolor[rgb]{ .867,  .922,  .969}   GPT-5-mini    & Live-SWE-Agent & 33.07\% & 30.97\% & 27.63\% & 27.42\% & 19.81\% & 23.85\% & 21.21\% & 23.26\% & 16.25\% & 36.36\% & 18.75\% & 20.00\% & 18.97\% & 15.09\% & 20.00\% \\
    \rowcolor[rgb]{ .867,  .922,  .969}       & \toolname{} & 33.20\% & 42.13\% & 37.89\% & 35.00\% & 42.86\% & 33.80\% & 32.81\% & 40.00\% & 33.33\% & 38.30\% & 34.88\%     & 32.61\%     & 45.71\%     & 43.59\%     & 42.11\% \\
    
    \bottomrule
    \end{tabular}}%
  \label{tab:sec_file_loc}%
\end{table*}%

%% file: tab/secbench_method_loc.tex
\begin{table*}[htbp]
  \centering
  \caption{Method-level localization accuracy on SEC-bench.}
    \resizebox{1.0\linewidth}{!}{\begin{tabular}{lcccccccccccccccc}
    \toprule
    \rowcolor[rgb]{ .906,  .902,  .902} \multicolumn{1}{l}{Model} & Method & Turn 1 & Turn 2 & Turn 3 & Turn 4 & Turn 5 & Turn 6 & Turn 7 & Turn 8 & Turn 9 & Turn 10 & Turn 11 & Turn 12 & Turn 13 & Turn 14 & Turn 15 \\
    \midrule
    
    \rowcolor[rgb]{ .886,  .937,  .855}       & Vanilla & 38.20\% & 34.62\% & 26.92\% & 21.43\% & 24.32\% & 20.45\% & 22.45\% & 21.43\% & 20.51\% & 43.75\% & 34.09\% & 20.45\% & 29.63\% & 21.88\% & 13.79\% \\
    \rowcolor[rgb]{ .886,  .937,  .855} DeepSeek-v3.1 & Live-SWE-Agent & 31.63\% & 29.03\% & 35.29\% & 31.48\% & 18.52\% & 41.94\% & 27.59\% & 28.21\% & 25.00\% & 26.92\% & 24.39\% & 17.86\% & 33.33\% & -     & - \\
    \rowcolor[rgb]{ .886,  .937,  .855}       & \toolname{} & 36.69\% & 29.25\% & 26.74\% & 28.92\% & 29.63\% & 25.76\% & 25.68\% & 28.57\% & 30.00\% & 23.53\% & 30.00\% & 20.37\% & -     & -     & - \\

    \rowcolor[rgb]{ .776,  .878,  .706}       & Vanilla & 40.20\% & 34.26\% & 39.44\% & 29.41\% & 31.91\% & -     & -     & -     & -     & -     & -     & -     & -     & -     & - \\
    \rowcolor[rgb]{ .776,  .878,  .706} DeepSeek-v3.2 & Live-SWE-Agent & 39.38\% & 33.88\% & 40.28\% & 38.81\% & 28.30\% & 28.57\% & 29.79\% & 30.77\% & 33.33\% & 22.22\% & 21.21\% & 23.33\% & 26.32\% & -     & - \\
    \rowcolor[rgb]{ .776,  .878,  .706}       & \toolname{} & 41.84\% & 36.52\% & 30.99\% & 38.71\% & 27.12\% & 31.71\% & 47.22\% & 21.21\% & 31.25\% & 23.08\% & -     & -     & -     & -     & - \\

    \rowcolor[rgb]{ 1,  .949,  .8}       & Vanilla & 41.94\% & 50.00\% & 42.31\% & 25.00\% & 33.33\% & 19.05\% & 38.10\% & 16.67\% & 40.00\% & 40.00\% & -     & -     & -     & -     & - \\
    \rowcolor[rgb]{ 1,  .949,  .8} Qwen3.5-Plus & Live-SWE-Agent & 52.08\% & 36.17\% & 50.00\% & 42.11\% & 40.00\% & 31.58\% & 53.85\% & 35.71\% & 27.27\% & 20.00\% & 42.86\% & 37.50\% & 38.46\% & 62.50\% & 37.50\% \\
    \rowcolor[rgb]{ 1,  .949,  .8}       & \toolname{} & 47.47\% & 48.28\% & 35.71\% & 36.67\% & 46.43\% & 34.78\% & 20.00\% & 13.33\% & 33.33\% & 25.00\% & 23.53\% & 20.00\% & 40.00\% & 25.00\% & - \\

    \rowcolor[rgb]{ .867,  .922,  .969}       & Vanilla & 22.62\% & 23.18\% & 14.18\% & 18.81\% & 17.31\% & 15.38\% & 13.95\% & 13.25\% & 12.99\% & 8.86\% & 12.99\% & 8.93\% & 14.04\% & 16.67\% & 12.00\% \\
    \rowcolor[rgb]{ .867,  .922,  .969} GPT-5-mini & Live-SWE-Agent & 28.97\% & 27.42\% & 22.63\% & 20.39\% & 12.24\% & 17.20\% & 17.07\% & 16.44\% & 13.04\% & 20.45\% & 6.67\% & 11.48\% & 9.80\% & 7.32\% & 11.36\% \\
    \rowcolor[rgb]{ .867,  .922,  .969}       & \toolname{} & 23.41\% & 33.55\% & 24.18\% & 34.21\% & 33.33\% & 20.63\% & 24.07\% & 25.00\% & 21.15\% & 16.67\% & 11.36\%     & 19.05\%     & 24.24\%     & 24.32\%     & 21.62\% \\

    \bottomrule
    \end{tabular}}%
  \label{tab:sec_method_loc}%
\end{table*}%

%% file: tab/transfer_result_full.tex
\begin{table*}[htbp]
  \centering
  \vspace{-1em}
  \caption{Experience transfer result on the VUL4J dataset.} 
  \vspace{-1em}
  \renewcommand{\arraystretch}{0.1}
  \resizebox{\textwidth}{!}{
    \small
    \begin{tabular}{lcccccc}
    \toprule 
    VUL & 
    \makecell{Qwen3.5-Plus Vanilla} & 
    \makecell{Qwen3.5-Plus Transfer} & 
    \makecell{Qwen3-Max Vanilla} & 
    \makecell{Qwen3-Max Transfer} & 
    \makecell{Qwen3-Coder-Plus Vanilla} & 
    \makecell{Qwen3-Coder-Plus Transfer} \\
    \midrule 
    VUL4J-1  & \XSolidBrush     & \XSolidBrush     & \XSolidBrush     & \XSolidBrush     & \XSolidBrush     & \XSolidBrush \\
    \rowcolor{lightgray}
    VUL4J-2  & \Checkmark     & \Checkmark     & \Checkmark     & \Checkmark     & \Checkmark     & \Checkmark \\
    VUL4J-3  & \XSolidBrush     & \XSolidBrush     & \XSolidBrush     & \XSolidBrush     & \XSolidBrush     & \XSolidBrush \\
    \rowcolor{lightgray}
    VUL4J-4  & \XSolidBrush     & \XSolidBrush     & \XSolidBrush     & \XSolidBrush     & \XSolidBrush     & \XSolidBrush \\
    VUL4J-5  & \XSolidBrush     & \XSolidBrush     & \XSolidBrush     & \XSolidBrush     & \XSolidBrush     & \XSolidBrush \\
    \rowcolor{lightgray}
    VUL4J-6  & \XSolidBrush     & \XSolidBrush     & \XSolidBrush     & \XSolidBrush     & \XSolidBrush     & \XSolidBrush \\
    VUL4J-7  & \XSolidBrush     & \XSolidBrush     & \XSolidBrush     & \XSolidBrush     & \XSolidBrush     & \XSolidBrush \\
    \rowcolor{lightgray}
    VUL4J-8  & \XSolidBrush     & \XSolidBrush     & \XSolidBrush     & \XSolidBrush     & \XSolidBrush     & \XSolidBrush \\
    VUL4J-9  & \Checkmark     & \Checkmark     & \Checkmark     & \Checkmark     & \Checkmark     & \Checkmark \\
    \rowcolor{lightgray}
    VUL4J-10 & \Checkmark     & \Checkmark     & \XSolidBrush     & \Checkmark     & \XSolidBrush     & \Checkmark \\
    VUL4J-11 & \Checkmark     & \Checkmark     & \Checkmark     & \Checkmark     & \Checkmark     & \Checkmark \\
    \rowcolor{lightgray}
    VUL4J-12 & \XSolidBrush     & \XSolidBrush     & \XSolidBrush     & \XSolidBrush     & \XSolidBrush     & \XSolidBrush \\
    VUL4J-13 & \XSolidBrush     & \XSolidBrush     & \XSolidBrush     & \XSolidBrush     & \XSolidBrush     & \XSolidBrush \\
    \rowcolor{lightgray}
    VUL4J-14 & \XSolidBrush     & \XSolidBrush     & \XSolidBrush     & \XSolidBrush     & \XSolidBrush     & \XSolidBrush \\
    VUL4J-15 & \XSolidBrush     & \XSolidBrush     & \XSolidBrush     & \XSolidBrush     & \XSolidBrush     & \XSolidBrush \\
    \rowcolor{lightgray}
    VUL4J-16 & \XSolidBrush     & \XSolidBrush     & \XSolidBrush     & \XSolidBrush     & \XSolidBrush     & \XSolidBrush \\
    VUL4J-17 & \XSolidBrush     & \XSolidBrush     & \XSolidBrush     & \XSolidBrush     & \XSolidBrush     & \XSolidBrush \\
    \rowcolor{lightgray}
    VUL4J-18 & \Checkmark     & \Checkmark     & \Checkmark     & \Checkmark     & \XSolidBrush     & \Checkmark \\
    VUL4J-19 & \Checkmark     & \Checkmark     & \XSolidBrush     & \XSolidBrush     & \XSolidBrush     & \Checkmark \\
    \rowcolor{lightgray}
    VUL4J-20 & \Checkmark     & \Checkmark     & \XSolidBrush     & \Checkmark     & \Checkmark     & \XSolidBrush \\
    VUL4J-21 & \XSolidBrush     & \XSolidBrush     & \XSolidBrush     & \XSolidBrush     & \XSolidBrush     & \XSolidBrush \\
    \rowcolor{lightgray}
    VUL4J-22 & \Checkmark     & \Checkmark     & \XSolidBrush     & \XSolidBrush     & \Checkmark     & \XSolidBrush \\
    VUL4J-23 & \XSolidBrush     & \XSolidBrush     & \XSolidBrush     & \XSolidBrush     & \XSolidBrush     & \XSolidBrush \\
    \rowcolor{lightgray}
    VUL4J-24 & \XSolidBrush     & \XSolidBrush     & \XSolidBrush     & \XSolidBrush     & \XSolidBrush     & \XSolidBrush \\
    VUL4J-25 & \Checkmark     & \Checkmark     & \Checkmark     & \XSolidBrush     & \XSolidBrush     & \XSolidBrush \\
    \rowcolor{lightgray}
    VUL4J-26 & \XSolidBrush     & \XSolidBrush     & \XSolidBrush     & \XSolidBrush     & \XSolidBrush     & \XSolidBrush \\
    VUL4J-27 & \XSolidBrush     & \XSolidBrush     & \XSolidBrush     & \XSolidBrush     & \XSolidBrush     & \XSolidBrush \\
    \rowcolor{lightgray}
    VUL4J-28 & \XSolidBrush     & \XSolidBrush     & \XSolidBrush     & \XSolidBrush     & \XSolidBrush     & \XSolidBrush \\
    VUL4J-29 & \XSolidBrush     & \XSolidBrush     & \XSolidBrush     & \XSolidBrush     & \XSolidBrush     & \XSolidBrush \\
    \rowcolor{lightgray}
    VUL4J-30 & \XSolidBrush     & \XSolidBrush     & \XSolidBrush     & \XSolidBrush     & \XSolidBrush     & \XSolidBrush \\
    VUL4J-31 & \XSolidBrush     & \XSolidBrush     & \XSolidBrush     & \XSolidBrush     & \XSolidBrush     & \XSolidBrush \\
    \rowcolor{lightgray}
    VUL4J-32 & \XSolidBrush     & \XSolidBrush     & \XSolidBrush     & \XSolidBrush     & \XSolidBrush     & \XSolidBrush \\
    VUL4J-33 & \XSolidBrush     & \XSolidBrush     & \XSolidBrush     & \XSolidBrush     & \XSolidBrush     & \XSolidBrush \\
    \rowcolor{lightgray}
    VUL4J-34 & \XSolidBrush     & \XSolidBrush     & \XSolidBrush     & \XSolidBrush     & \XSolidBrush     & \XSolidBrush \\
    VUL4J-35 & \XSolidBrush     & \XSolidBrush     & \XSolidBrush     & \XSolidBrush     & \XSolidBrush     & \XSolidBrush \\
    \rowcolor{lightgray}
    VUL4J-36 & \Checkmark     & \Checkmark     & \XSolidBrush     & \XSolidBrush     & \XSolidBrush     & \Checkmark \\
    VUL4J-37 & \XSolidBrush     & \XSolidBrush     & \XSolidBrush     & \XSolidBrush     & \XSolidBrush     & \XSolidBrush \\
    \rowcolor{lightgray}
    VUL4J-38 & \Checkmark     & \Checkmark     & \Checkmark     & \Checkmark     & \XSolidBrush     & \XSolidBrush \\
    VUL4J-39 & \XSolidBrush     & \XSolidBrush     & \XSolidBrush     & \XSolidBrush     & \XSolidBrush     & \XSolidBrush \\
    \rowcolor{lightgray}
    VUL4J-40 & \XSolidBrush     & \XSolidBrush     & \XSolidBrush     & \XSolidBrush     & \XSolidBrush     & \XSolidBrush \\
    VUL4J-41 & \Checkmark     & \Checkmark     & \Checkmark     & \Checkmark     & \XSolidBrush     & \XSolidBrush \\
    \rowcolor{lightgray}
    VUL4J-42 & \XSolidBrush     & \XSolidBrush     & \XSolidBrush     & \XSolidBrush     & \XSolidBrush     & \XSolidBrush \\
    VUL4J-43 & \Checkmark     & \Checkmark     & \Checkmark     & \Checkmark     & \Checkmark     & \Checkmark \\
    \rowcolor{lightgray}
    VUL4J-44 & \XSolidBrush     & \XSolidBrush     & \XSolidBrush     & \XSolidBrush     & \XSolidBrush     & \XSolidBrush \\
    VUL4J-45 & \Checkmark     & \Checkmark     & \Checkmark     & \Checkmark     & \Checkmark     & \Checkmark \\
    \rowcolor{lightgray}
    VUL4J-46 & \XSolidBrush     & \XSolidBrush     & \XSolidBrush     & \XSolidBrush     & \XSolidBrush     & \XSolidBrush \\
    VUL4J-47 & \XSolidBrush     & \XSolidBrush     & \XSolidBrush     & \XSolidBrush     & \XSolidBrush     & \XSolidBrush \\
    \rowcolor{lightgray}
    VUL4J-48 & \Checkmark     & \Checkmark     & \Checkmark     & \Checkmark     & \Checkmark     & \Checkmark \\
    VUL4J-49 & \Checkmark     & \Checkmark     & \Checkmark     & \Checkmark     & \Checkmark     & \Checkmark \\
    \rowcolor{lightgray}
    VUL4J-50 & \XSolidBrush     & \XSolidBrush     & \XSolidBrush     & \XSolidBrush     & \XSolidBrush     & \XSolidBrush \\
    VUL4J-51 & \XSolidBrush     & \XSolidBrush     & \XSolidBrush     & \XSolidBrush     & \XSolidBrush     & \XSolidBrush \\
    \rowcolor{lightgray}
    VUL4J-52 & \Checkmark     & \Checkmark     & \Checkmark     & \XSolidBrush     & \XSolidBrush     & \Checkmark \\
    VUL4J-53 & \XSolidBrush     & \XSolidBrush     & \XSolidBrush     & \XSolidBrush     & \XSolidBrush     & \XSolidBrush \\
    \rowcolor{lightgray}
    VUL4J-54 & \XSolidBrush     & \XSolidBrush     & \XSolidBrush     & \XSolidBrush     & \XSolidBrush     & \XSolidBrush \\
    VUL4J-55 & \XSolidBrush     & \XSolidBrush     & \XSolidBrush     & \XSolidBrush     & \XSolidBrush     & \XSolidBrush \\
    \rowcolor{lightgray}
    VUL4J-56 & \XSolidBrush     & \Checkmark     & \XSolidBrush     & \XSolidBrush     & \Checkmark     & \Checkmark \\
    VUL4J-57 & \Checkmark     & \Checkmark     & \XSolidBrush     & \Checkmark     & \Checkmark     & \XSolidBrush \\
    \rowcolor{lightgray}
    VUL4J-58 & \XSolidBrush     & \Checkmark     & \XSolidBrush     & \XSolidBrush     & \XSolidBrush     & \XSolidBrush \\
    VUL4J-59 & \Checkmark     & \Checkmark     & \Checkmark     & \Checkmark     & \XSolidBrush     & \XSolidBrush \\
    \rowcolor{lightgray}
    VUL4J-60 & \Checkmark     & \Checkmark     & \XSolidBrush     & \XSolidBrush     & \Checkmark     & \Checkmark \\
    VUL4J-61 & \Checkmark     & \Checkmark     & \Checkmark     & \Checkmark     & \Checkmark     & \Checkmark \\
    \rowcolor{lightgray}
    VUL4J-62 & \Checkmark     & \Checkmark     & \Checkmark     & \Checkmark     & \Checkmark     & \Checkmark \\
    VUL4J-63 & \XSolidBrush     & \XSolidBrush     & \XSolidBrush     & \XSolidBrush     & \XSolidBrush     & \XSolidBrush \\
    \rowcolor{lightgray}
    VUL4J-64 & \Checkmark     & \Checkmark     & \Checkmark     & \Checkmark     & \Checkmark     & \Checkmark \\
    VUL4J-65 & \Checkmark     & \Checkmark     & \Checkmark     & \Checkmark     & \Checkmark     & \Checkmark \\
    \rowcolor{lightgray}
    VUL4J-66 & \Checkmark     & \Checkmark     & \Checkmark     & \Checkmark     & \XSolidBrush     & \XSolidBrush \\
    VUL4J-67 & \XSolidBrush     & \XSolidBrush     & \XSolidBrush     & \XSolidBrush     & \XSolidBrush     & \XSolidBrush \\
    \rowcolor{lightgray}
    VUL4J-68 & \Checkmark     & \Checkmark     & \Checkmark     & \Checkmark     & \Checkmark     & \Checkmark \\
    VUL4J-69 & \Checkmark     & \Checkmark     & \Checkmark     & \Checkmark     & \Checkmark     & \Checkmark \\
    \rowcolor{lightgray}
    VUL4J-70 & \Checkmark     & \Checkmark     & \Checkmark     & \Checkmark     & \Checkmark     & \XSolidBrush \\
    VUL4J-71 & \XSolidBrush     & \Checkmark     & \Checkmark     & \Checkmark     & \Checkmark     & \Checkmark \\
    \rowcolor{lightgray}
    VUL4J-72 & \XSolidBrush     & \XSolidBrush     & \XSolidBrush     & \XSolidBrush     & \XSolidBrush     & \XSolidBrush \\
    VUL4J-73 & \XSolidBrush     & \XSolidBrush     & \XSolidBrush     & \XSolidBrush     & \XSolidBrush     & \XSolidBrush \\
    \rowcolor{lightgray}
    VUL4J-74 & \XSolidBrush     & \XSolidBrush     & \XSolidBrush     & \XSolidBrush     & \XSolidBrush     & \XSolidBrush \\
    VUL4J-75 & \XSolidBrush     & \XSolidBrush     & \XSolidBrush     & \XSolidBrush     & \XSolidBrush     & \XSolidBrush \\
    \rowcolor{lightgray}
    VUL4J-76 & \Checkmark     & \Checkmark     & \XSolidBrush     & \XSolidBrush     & \XSolidBrush     & \XSolidBrush \\
    VUL4J-77 & \Checkmark     & \Checkmark     & \Checkmark     & \Checkmark     & \Checkmark     & \Checkmark \\
    \rowcolor{lightgray}
    VUL4J-78 & \XSolidBrush     & \XSolidBrush     & \XSolidBrush     & \XSolidBrush     & \XSolidBrush     & \XSolidBrush \\
    VUL4J-79 & \Checkmark     & \Checkmark     & \XSolidBrush     & \Checkmark     & \Checkmark     & \Checkmark \\
    \bottomrule 
    \end{tabular}%
  }
  \label{tab:transfer_result_full}
\end{table*}%

%% file: tab/early_stop_detail.tex
\begin{table*}[htbp]
  \centering
  \renewcommand{\arraystretch}{0.69} 
  \caption{Breakdown of new fixes and additional costs per turn on PATCHEVAL, used as a criterion for the early-stopping strategy.}
    \resizebox{1.0\linewidth}{!}{
    \footnotesize
    \begin{tabular}{lrrrrrrrrr}
    
    \toprule
    Model & Turn  & CVEs  & New Fixes & Cumulative Fixes & New Cost (\$) & Cumulative Cost (\$) & $\beta$ & $\gamma$ & $\alpha$  \\
    \midrule
    \multirow{15}[2]{*}{DeepSeek-v3.1} & 1     & 230   & 54    & 54    & 75.8867 & 75.8867 & $\infty$     & $\infty$     & $\infty$ \\
          & 2     & 176   & 28    & 82    & 64.7944 & 140.6811 & 0.5185 & 0.8538 & 0.6073 \\
          & 3     & 148   & 9     & 91    & 59.5065 & 200.1876 & 0.1098 & 0.423 & 0.2595 \\
          & 4     & 139   & 13    & 104   & 51.6457 & 251.8333 & 0.1429 & 0.258 & 0.5537 \\
          & 5     & 126   & 18    & 122   & 47.594 & 299.4273 & 0.1731 & 0.189 & 0.9158 \\
          & 6     & 108   & 11    & 133   & 41.7326 & 341.1599 & 0.0902 & 0.1394 & 0.6469 \\
          & 7     & 97    & 8     & 141   & 37.4431 & 378.603 & 0.0602 & 0.1098 & 0.5481 \\
          & 8     & 89    & 5     & 146   & 36.1995 & 414.8025 & 0.0355 & 0.0956 & 0.3709 \\
          & 9     & 84    & 3     & 149   & 30.5494 & 445.352 & 0.0205 & 0.0736 & 0.279 \\
          & 10    & 81    & 1     & 150   & 29.284 & 474.6359 & 0.0067 & 0.0658 & 0.1021 \\
          & 11    & 80    & 4     & 154   & 32.6795 & 507.3154 & 0.0267 & 0.0689 & 0.3873 \\
          & 12    & 76    & 4     & 158   & 30.8755 & 538.191 & 0.026 & 0.0609 & 0.4268 \\
          & 13    & 72    & 2     & 160   & 26.253 & 564.444 & 0.0127 & 0.0488 & 0.2595 \\
          & 14    & 70    & 3     & 163   & 30.2974 & 594.7414 & 0.0187 & 0.0537 & 0.3493 \\
          & 15    & -     & -     & -     & -     & -     & -     & -     & - \\
    \midrule
    \multirow{15}[2]{*}{Qwen3.5-Plus} & 1     & 230   & 131   & 131   & 137.3843 & 137.3843 & $\infty$     & $\infty$     & $\infty$ \\
          & 2     & 99    & 53    & 184   & 68.6887 & 206.073 & 0.4046 & 0.5   & 0.8092 \\
          & 3     & 46    & 15    & 199   & 40.2869 & 246.36 & 0.0815 & 0.1955 & 0.417 \\
          & 4     & 31    & 6     & 205   & 28.0739 & 274.4339 & 0.0302 & 0.114 & 0.2646 \\
          & 5     & 25    & 2     & 207   & 20.4882 & 294.9221 & 0.0098 & 0.0747 & 0.1307 \\
          & 6     & 23    & 2     & 209   & 22.8841 & 317.8062 & 0.0097 & 0.0776 & 0.1245 \\
          & 7     & 21    & 3     & 212   & 17.7669 & 335.5731 & 0.0144 & 0.0559 & 0.2568 \\
          & 8     & 18    & 2     & 214   & 18.2184 & 353.7915 & 0.0094 & 0.0543 & 0.1738 \\
          & 9     & 16    & 4     & 218   & 11.7496 & 365.5411 & 0.0187 & 0.0332 & 0.5628 \\
          & 10    & 12    & 1     & 219   & 12.4055 & 377.9466 & 0.0046 & 0.0339 & 0.1352 \\
          & 11    & 11    & 1     & 220   & 10.1384 & 388.0851 & 0.0046 & 0.0268 & 0.1702 \\
          & 12    & 10    & 0     & 220   & 10.2706 & 398.3557 & 0     & 0.0265 & 0 \\
          & 13    & -     & -     & -     & -     & -     & -     & -     & - \\
          & 14    & -     & -     & -     & -     & -     & -     & -     & - \\
          & 15    & -     & -     & -     & -     & -     & -     & -     & - \\
    \midrule
    \multirow{15}[2]{*}{Devstral-123B} & 1     & 230   & 124   & 124   & 104.8407 & 104.8407 & $\infty$     & $\infty$     & $\infty$ \\
          & 2     & 106   & 40    & 164   & 53.7326 & 158.5732 & 0.3226 & 0.5125 & 0.6294 \\
          & 3     & 66    & 15    & 179   & 36.4763 & 195.0495 & 0.0915 & 0.23  & 0.3976 \\
          & 4     & 51    & 13    & 192   & 26.5779 & 221.6274 & 0.0726 & 0.1363 & 0.533 \\
          & 5     & 38    & 3     & 195   & 24.9846 & 246.612 & 0.0156 & 0.1127 & 0.1386 \\
          & 6     & 35    & 6     & 201   & 19.3673 & 265.9793 & 0.0308 & 0.0785 & 0.3918 \\
          & 7     & 29    & 4     & 205   & 18.6482 & 284.6275 & 0.0199 & 0.0701 & 0.2838 \\
          & 8     & 25    & 2     & 207   & 15.5734 & 300.2009 & 0.0098 & 0.0547 & 0.1783 \\
          & 9     & 23    & 1     & 208   & 14.7502 & 314.9511 & 0.0048 & 0.0491 & 0.0983 \\
          & 10    & 22    & 2     & 210   & 11.872 & 326.8231 & 0.0096 & 0.0377 & 0.2551 \\
          & 11    & 20    & 1     & 211   & 8.8812 & 335.7044 & 0.0048 & 0.0272 & 0.1752 \\
          & 12    & 19    & 2     & 213   & 8.6617 & 344.3661 & 0.0095 & 0.0258 & 0.3674 \\
          & 13    & 17    & 2     & 215   & 10.5507 & 354.9168 & 0.0094 & 0.0306 & 0.3065 \\
          & 14    & 15    & 1     & 216   & 8.9027 & 363.8195 & 0.0047 & 0.0251 & 0.1854 \\
          & 15    & -     & -     & -     & -     & -     & -     & -     & - \\
    \midrule
    \multirow{15}[2]{*}{Devstral-24B} & 1     & 230   & 113   & 113   & 28.4529 & 28.4529 & $\infty$     & $\infty$     & $\infty$ \\
          & 2     & 117   & 49    & 162   & 18.1357 & 46.5886 & 0.4336 & 0.6374 & 0.6803 \\
          & 3     & 68    & 17    & 179   & 11.6543 & 58.243 & 0.1049 & 0.2502 & 0.4195 \\
          & 4     & 51    & 12    & 191   & 9.8368 & 68.0797 & 0.067 & 0.1689 & 0.3969 \\
          & 5     & 39    & 4     & 195   & 7.7828 & 75.8625 & 0.0209 & 0.1143 & 0.1832 \\
          & 6     & 35    & 4     & 199   & 6.9982 & 82.8607 & 0.0205 & 0.0922 & 0.2224 \\
          & 7     & 31    & 3     & 202   & 6.1207 & 88.9814 & 0.0151 & 0.0739 & 0.2041 \\
          & 8     & 28    & 5     & 207   & 5.8086 & 94.79 & 0.0248 & 0.0653 & 0.3792 \\
          & 9     & 23    & 6     & 213   & 3.9927 & 98.7828 & 0.029 & 0.0421 & 0.6881 \\
          & 10    & 17    & 1     & 214   & 3.4361 & 102.2188 & 0.0047 & 0.0348 & 0.135 \\
          & 11    & 16    & 2     & 216   & 2.7336 & 104.9524 & 0.0093 & 0.0267 & 0.3495 \\
          & 12    & 14    & 1     & 217   & 2.6001 & 107.5525 & 0.0046 & 0.0248 & 0.1869 \\
          & 13    & 13    & 1     & 218   & 2.4921 & 110.0446 & 0.0046 & 0.0232 & 0.1989 \\
          & 14    & 12    & 0     & 218   & 2.4103 & 112.4548 & 0     & 0.0219 & 0 \\
          & 15    & -     & -     & -     & -     & -     & -     & -     & - \\
    \midrule
    \multirow{15}[2]{*}{GPT-5-mini} & 1     & 230   & 105   & 105   & 41.1859 & 41.1859 & $\infty$     & $\infty$     & $\infty$ \\
          & 2     & 125   & 49    & 154   & 28.9068 & 70.0927 & 0.4666 & 0.7019 & 0.6648 \\
          & 3     & 76    & 22    & 176   & 17.4825 & 87.5752 & 0.1429 & 0.2494 & 0.5728 \\
          & 4     & 54    & 12    & 188   & 12.1541 & 99.7293 & 0.0682 & 0.1388 & 0.4913 \\
          & 5     & 42    & 7     & 195   & 12.2516 & 111.9809 & 0.0372 & 0.1228 & 0.3031 \\
          & 6     & 35    & 3     & 198   & 8.5106 & 120.4915 & 0.0154 & 0.076 & 0.2024 \\
          & 7     & 32    & 4     & 202   & 12.4333 & 132.9248 & 0.0202 & 0.1032 & 0.1958 \\
          & 8     & 28    & 2     & 204   & 14.608 & 147.5327 & 0.0099 & 0.1099 & 0.0901 \\
          & 9     & 26    & 2     & 206   & 9.4343 & 156.967 & 0.0098 & 0.0639 & 0.1533 \\
          & 10    & 24    & 4     & 210   & 9.369 & 166.336 & 0.0194 & 0.0597 & 0.3253 \\
          & 11    & 20    & 1     & 211   & 7.0953 & 173.4313 & 0.0048 & 0.0427 & 0.1116 \\
          & 12    & 19    & 1     & 212   & 9.1991 & 182.6304 & 0.0047 & 0.053 & 0.0894 \\
          & 13    & 18    & 2     & 214   & 4.9421 & 187.5725 & 0.0094 & 0.0271 & 0.3486 \\
          & 14    & 16    & 1     & 215   & 6.8364 & 194.4089 & 0.0047 & 0.0364 & 0.1282 \\
          & 15    & -     & -     & -     & -     & -     & -     & -     & - \\
    \bottomrule
    \end{tabular}}%
  \label{tab:early_stop_detail}%
\end{table*}%

%% file: reference.bib
@inproceedings{fu2022vulrepair,
  title={VulRepair: a T5-based automated software vulnerability repair},
  author={Fu, Michael and Tantithamthavorn, Chakkrit and Le, Trung and Nguyen, Van and Phung, Dinh},
  booktitle={Proceedings of the 30th ACM joint european software engineering conference and symposium on the foundations of software engineering},
  pages={935--947},
  year={2022}
}

@article{zhang2023pre,
  title={Pre-trained model-based automated software vulnerability repair: How far are we?},
  author={Zhang, Quanjun and Fang, Chunrong and Yu, Bowen and Sun, Weisong and Zhang, Tongke and Chen, Zhenyu},
  journal={IEEE Transactions on Dependable and Secure Computing},
  volume={21},
  number={4},
  pages={2507--2525},
  year={2023},
  publisher={IEEE}
}

@article{shen2020survey,
  title={A survey of automatic software vulnerability detection, program repair, and defect prediction techniques},
  author={Shen, Zhidong and Chen, Si},
  journal={Security and Communication Networks},
  volume={2020},
  number={1},
  pages={8858010},
  year={2020},
  publisher={Wiley Online Library}
}

@inproceedings{gao2019crash,
  title={Crash-avoiding program repair},
  author={Gao, Xiang and Mechtaev, Sergey and Roychoudhury, Abhik},
  booktitle={Proceedings of the 28th ACM SIGSOFT International Symposium on Software Testing and Analysis},
  pages={8--18},
  year={2019}
}

@article{gao2021beyond,
  title={Beyond tests: Program vulnerability repair via crash constraint extraction},
  author={Gao, Xiang and Wang, Bo and Duck, Gregory J and Ji, Ruyi and Xiong, Yingfei and Roychoudhury, Abhik},
  journal={ACM Transactions on Software Engineering and Methodology (TOSEM)},
  volume={30},
  number={2},
  pages={1--27},
  year={2021},
  publisher={ACM New York, NY, USA}
}

@inproceedings{fu2023toward,
  title={Toward more effective deep learning-based automated software vulnerability prediction, classification, and repair},
  author={Fu, Michael},
  booktitle={2023 IEEE/ACM 45th International Conference on Software Engineering: Companion Proceedings (ICSE-Companion)},
  pages={208--212},
  year={2023},
  organization={IEEE}
}

@inproceedings{zhou2024out,
  title={Out of sight, out of mind: Better automatic vulnerability repair by broadening input ranges and sources},
  author={Zhou, Xin and Kim, Kisub and Xu, Bowen and Han, DongGyun and Lo, David},
  booktitle={Proceedings of the IEEE/ACM 46th international conference on software engineering},
  pages={1--13},
  year={2024}
}

@inproceedings{costin2024evaluating,
  title={Evaluating zero-shot Chatgpt performance on predicting CVE data from vulnerability descriptions},
  author={Costin, Andrei and Turtiainen, Hannu and Yousefnezhad, Narges and Bogulean, Vadim and H{\"a}m{\"a}l{\"a}inen, Timo},
  booktitle={Proceedings of the European Conference on Cyber Warfare and Security},
  number={1},
  year={2024},
  organization={Academic Conferences International Ltd}
}

@inproceedings{bao2025smart,
  title={A Smart Contract Vulnerability Detection Method Based on Graph Neural Networks and Zero-Shot Learning},
  author={Bao, Kunhang and Chen, Shanshan},
  booktitle={International Conference on Blockchain and Trustworthy Systems},
  pages={32--46},
  year={2025},
  organization={Springer}
}

@article{zhou2025large,
  title={Large language model for vulnerability detection and repair: Literature review and the road ahead},
  author={Zhou, Xin and Cao, Sicong and Sun, Xiaobing and Lo, David},
  journal={ACM Transactions on Software Engineering and Methodology},
  volume={34},
  number={5},
  pages={1--31},
  year={2025},
  publisher={ACM New York, NY}
}

@article{zhang2026vulnresolver,
  title={VulnResolver: A Hybrid Agent Framework for LLM-Based Automated Vulnerability Issue Resolution},
  author={Zhang, Mingming and Wang, Xu and Zhang, Jian and Meng, Xiangxin and Zhang, Jiayi and Hu, Chunming},
  journal={arXiv preprint arXiv:2601.13933},
  year={2026}
}

@inproceedings{xia2024automated,
  title={Automated program repair via conversation: Fixing 162 out of 337 bugs for \$0.42 each using chatgpt},
  author={Xia, Chunqiu Steven and Zhang, Lingming},
  booktitle={Proceedings of the 33rd ACM SIGSOFT International Symposium on Software Testing and Analysis},
  pages={819--831},
  year={2024}
}

@article{ye2025well,
  title={Well Begun is Half Done: Location-Aware and Trace-Guided Iterative Automated Vulnerability Repair},
  author={Ye, Zhenlei and Sun, Xiaobing and Cao, Sicong and Bo, Lili and Li, Bin},
  journal={arXiv preprint arXiv:2512.20203},
  year={2025}
}

@inproceedings{fakih2025llm4cve,
  title={Llm4cve: Enabling iterative automated vulnerability repair with large language models},
  author={Fakih, Mohamad and Dharmaji, Rahul and Bouzidi, Halima and Araya, Gustavo Quiros and Ogundare, Oluwatosin and Siddika, Mst Ayesha and Al Faruque, Mohammad Abdullah},
  booktitle={2025 28th Euromicro Conference on Digital System Design (DSD)},
  pages={592--599},
  year={2025},
  organization={IEEE}
}

@article{wei2025patcheval,
  title={PATCHEVAL: A New Benchmark for Evaluating LLMs on Patching Real-World Vulnerabilities},
  author={Wei, Zichao and Zeng, Jun and Wen, Ming and Yu, Zeliang and Cheng, Kai and Zhu, Yiding and Guo, Jingyi and Zhou, Shiqi and Yin, Le and Su, Xiaodong and others},
  journal={arXiv preprint arXiv:2511.11019},
  year={2025}
}

@article{lee2025sec,
  title={Sec-bench: Automated benchmarking of llm agents on real-world software security tasks},
  author={Lee, Hwiwon and Zhang, Ziqi and Lu, Hanxiao and Zhang, Lingming},
  journal={arXiv preprint arXiv:2506.11791},
  year={2025}
}

@article{yang2024swe,
  title={Swe-agent: Agent-computer interfaces enable automated software engineering},
  author={Yang, John and Jimenez, Carlos E and Wettig, Alexander and Lieret, Kilian and Yao, Shunyu and Narasimhan, Karthik and Press, Ofir},
  journal={Advances in Neural Information Processing Systems},
  volume={37},
  pages={50528--50652},
  year={2024}
}

@inproceedings{DBLP:conf/icse/NollerS0R22,
  author       = {Yannic Noller and
                  Ridwan Shariffdeen and
                  Xiang Gao and
                  Abhik Roychoudhury},
  title        = {Trust Enhancement Issues in Program Repair},
  booktitle    = {44th {IEEE/ACM} 44th International Conference on Software Engineering,
                  {ICSE} 2022, Pittsburgh, PA, USA, May 25-27, 2022},
  pages        = {2228--2240},
  publisher    = {{ACM}},
  year         = {2022},
  url          = {https://doi.org/10.1145/3510003.3510040},
  doi          = {10.1145/3510003.3510040},
  bibsource    = {dblp computer science bibliography, https://dblp.org}
}

@article{DBLP:journals/tdsc/BellevilleSVAF21,
  author       = {Brian Belleville and
                  Wenbo Shen and
                  Stijn Volckaert and
                  Ahmed M. Azab and
                  Michael Franz},
  title        = {{KALD:} Detecting Direct Pointer Disclosure Vulnerabilities},
  journal      = {{IEEE} Trans. Dependable Secur. Comput.},
  volume       = {18},
  number       = {3},
  pages        = {1369--1377},
  year         = {2021},
  url          = {https://doi.org/10.1109/TDSC.2019.2915829},
  doi          = {10.1109/TDSC.2019.2915829},
  bibsource    = {dblp computer science bibliography, https://dblp.org}
}

@misc{spotbugs,
title = {SpotBugs: Find bugs in Java Programs},
publisher = {\url{https://spotbugs.github.io/}},
year = {2022},
note = {Last accessed: November 20, 2022}
}

@misc{infer,
title = {Infer Static Analyzer},
publisher = {\url{https://fbinfer.com/}},
year = {2022},
note = {Last accessed: November 20, 2022}
}

@article{DBLP:journals/tse/ChenKM23,
  author       = {Zimin Chen and
                  Steve Kommrusch and
                  Martin Monperrus},
  title        = {Neural Transfer Learning for Repairing Security Vulnerabilities in
                  {C} Code},
  journal      = {{IEEE} Trans. Software Eng.},
  volume       = {49},
  number       = {1},
  pages        = {147--165},
  year         = {2023},
  url          = {https://doi.org/10.1109/TSE.2022.3147265},
  doi          = {10.1109/TSE.2022.3147265},
  bibsource    = {dblp computer science bibliography, https://dblp.org}
}

@article{DBLP:journals/corr/abs-2203-05166,
  author       = {Quanjun Zhang and
                  Yuan Zhao and
                  Weisong Sun and
                  Chunrong Fang and
                  Ziyuan Wang and
                  Lingming Zhang},
  title        = {Program Repair: Automated vs. Manual},
  journal      = {CoRR},
  volume       = {abs/2203.05166},
  year         = {2022},
  url          = {https://doi.org/10.48550/arXiv.2203.05166},
  doi          = {10.48550/ARXIV.2203.05166},
  eprinttype    = {arXiv},
  eprint       = {2203.05166},
  bibsource    = {dblp computer science bibliography, https://dblp.org}
}

@article{DBLP:journals/tse/ChiQLZY23,
  author       = {Jianlei Chi and
                  Yu Qu and
                  Ting Liu and
                  Qinghua Zheng and
                  Heng Yin},
  title        = {SeqTrans: Automatic Vulnerability Fix Via Sequence to Sequence Learning},
  journal      = {{IEEE} Trans. Software Eng.},
  volume       = {49},
  number       = {2},
  pages        = {564--585},
  year         = {2023},
  url          = {https://doi.org/10.1109/TSE.2022.3156637},
  doi          = {10.1109/TSE.2022.3156637},
  bibsource    = {dblp computer science bibliography, https://dblp.org}
}

@article{DBLP:journals/corr/abs-2107-08760,
  author       = {Guru Prasad Bhandari and
                  Amara Naseer and
                  Leon Moonen},
  title        = {CVEfixes: Automated Collection of Vulnerabilities and Their Fixes
                  from Open-Source Software},
  journal      = {CoRR},
  volume       = {abs/2107.08760},
  year         = {2021},
  url          = {https://arxiv.org/abs/2107.08760},
  eprinttype    = {arXiv},
  eprint       = {2107.08760},
  bibsource    = {dblp computer science bibliography, https://dblp.org}
}

@inproceedings{DBLP:conf/kbse/WenLYGY25,
  author       = {Xin{-}Cheng Wen and
                  Zirui Lin and
                  Yijun Yang and
                  Cuiyun Gao and
                  Deheng Ye},
  title        = {Vul-R2: {A} Reasoning {LLM} for Automated Vulnerability Repair},
  booktitle    = {40th {IEEE/ACM} International Conference on Automated Software Engineering,
                  {ASE} 2025, Seoul, Korea, Republic of, November 16-20, 2025},
  pages        = {26--38},
  publisher    = {{IEEE}},
  year         = {2025},
  url          = {https://doi.org/10.1109/ASE63991.2025.00011},
  doi          = {10.1109/ASE63991.2025.00011},
  bibsource    = {dblp computer science bibliography, https://dblp.org}
}

@article{DBLP:journals/tosem/ZhouCSL25,
  author       = {Xin Zhou and
                  Sicong Cao and
                  Xiaobing Sun and
                  David Lo},
  title        = {Large Language Model for Vulnerability Detection and Repair: Literature
                  Review and the Road Ahead},
  journal      = {{ACM} Trans. Softw. Eng. Methodol.},
  volume       = {34},
  number       = {5},
  pages        = {145:1--145:31},
  year         = {2025},
  url          = {https://doi.org/10.1145/3708522},
  doi          = {10.1145/3708522},
  bibsource    = {dblp computer science bibliography, https://dblp.org}
}

@article{DBLP:journals/corr/abs-2504-07634,
  author       = {Zhengyao Liu and
                  Yunlong Ma and
                  Jingxuan Xu and
                  Junchen Ai and
                  Xiang Gao and
                  Hailong Sun and
                  Abhik Roychoudhury},
  title        = {Agent That Debugs: Dynamic State-Guided Vulnerability Repair},
  journal      = {CoRR},
  volume       = {abs/2504.07634},
  year         = {2025},
  url          = {https://doi.org/10.48550/arXiv.2504.07634},
  doi          = {10.48550/ARXIV.2504.07634},
  eprinttype    = {arXiv},
  eprint       = {2504.07634},
  bibsource    = {dblp computer science bibliography, https://dblp.org}
}

@article{DBLP:journals/corr/abs-2601-13933,
  author       = {Mingming Zhang and
                  Xu Wang and
                  Jian Zhang and
                  Xiangxin Meng and
                  Jiayi Zhang and
                  Chunming Hu},
  title        = {VulnResolver: {A} Hybrid Agent Framework for LLM-Based Automated Vulnerability
                  Issue Resolution},
  journal      = {CoRR},
  volume       = {abs/2601.13933},
  year         = {2026},
  url          = {https://doi.org/10.48550/arXiv.2601.13933},
  doi          = {10.48550/ARXIV.2601.13933},
  eprinttype    = {arXiv},
  eprint       = {2601.13933},
  bibsource    = {dblp computer science bibliography, https://dblp.org}
}

@article{kim2026patchislandorchestrationllmagents,
      title={PatchIsland: Orchestration of LLM Agents for Continuous Vulnerability Repair}, 
      author={Wonyoung Kim and Seunggi Min and Minjae Gwon and Dowoo Baik and Haein Lee and Hyeon Heo and Minjae Lee and Min Woo Baek and Yonghwi Jin and Younggi Park and Yunjae Choi and Taesoo Kim and Sangdon Park and Insu Yun},
      year={2026},
      journal={arXiv preprint arXiv:2601.17471},
}

@inproceedings{DBLP:conf/dsd/FakihDBAOSF25,
  author       = {Mohamad Fakih and
                  Rahul Dharmaji and
                  Halima Bouzidi and
                  Gustavo Quiros Araya and
                  Oluwatosin Ogundare and
                  Mst{-}Ayesha Siddika and
                  Mohammad Abdullah Al Faruque},
  title        = {{LLM4CVE:} Enabling Iterative Automated Vulnerability Repair with
                  Large Language Models},
  booktitle    = {28th Euromicro Conference on Digital System Design, {DSD} 2025, Salerno,
                  Italy, September 10-12, 2025},
  pages        = {592--599},
  publisher    = {{IEEE}},
  year         = {2025},
  url          = {https://doi.org/10.1109/DSD67783.2025.00087},
  doi          = {10.1109/DSD67783.2025.00087},
  bibsource    = {dblp computer science bibliography, https://dblp.org}
}

@article{DBLP:journals/corr/abs-2509-03331,
  author       = {Weizhe Wang and
                  Wei Ma and
                  Qiang Hu and
                  Yao Zhang and
                  Jianfei Sun and
                  Bin Wu and
                  Yang Liu and
                  Guangquan Xu and
                  Lingxiao Jiang},
  title        = {VulnRepairEval: An Exploit-Based Evaluation Framework for Assessing
                  Large Language Model Vulnerability Repair Capabilities},
  journal      = {CoRR},
  volume       = {abs/2509.03331},
  year         = {2025},
  url          = {https://doi.org/10.48550/arXiv.2509.03331},
  doi          = {10.48550/ARXIV.2509.03331},
  eprinttype    = {arXiv},
  eprint       = {2509.03331},
  bibsource    = {dblp computer science bibliography, https://dblp.org}
}

@article{wang2024openhands,
  title={Openhands: An open platform for ai software developers as generalist agents},
  author={Wang, Xingyao and Li, Boxuan and Song, Yufan and Xu, Frank F and Tang, Xiangru and Zhuge, Mingchen and Pan, Jiayi and Song, Yueqi and Li, Bowen and Singh, Jaskirat and others},
  journal={arXiv preprint arXiv:2407.16741},
  year={2024}
}

@inproceedings{zhao2024expel,
  title={Expel: Llm agents are experiential learners},
  author={Zhao, Andrew and Huang, Daniel and Xu, Quentin and Lin, Matthieu and Liu, Yong-Jin and Huang, Gao},
  booktitle={Proceedings of the AAAI Conference on Artificial Intelligence},
  volume={38},
  number={17},
  pages={19632--19642},
  year={2024}
}

@article{wu2025evolver,
  title={Evolver: Self-evolving llm agents through an experience-driven lifecycle},
  author={Wu, Rong and Wang, Xiaoman and Mei, Jianbiao and Cai, Pinlong and Fu, Daocheng and Yang, Cheng and Wen, Licheng and Yang, Xuemeng and Shen, Yufan and Wang, Yuxin and others},
  journal={arXiv preprint arXiv:2510.16079},
  year={2025}
}

@article{zhao2025absolute,
  title={Absolute zero: Reinforced self-play reasoning with zero data},
  author={Zhao, Andrew and Wu, Yiran and Yue, Yang and Wu, Tong and Xu, Quentin and Lin, Matthieu and Wang, Shenzhi and Wu, Qingyun and Zheng, Zilong and Huang, Gao},
  journal={arXiv preprint arXiv:2505.03335},
  year={2025}
}

@article{huang2025r,
  title={R-zero: Self-evolving reasoning llm from zero data},
  author={Huang, Chengsong and Yu, Wenhao and Wang, Xiaoyang and Zhang, Hongming and Li, Zongxia and Li, Ruosen and Huang, Jiaxin and Mi, Haitao and Yu, Dong},
  journal={arXiv preprint arXiv:2508.05004},
  year={2025}
}

@article{chen2025multi,
  title={Multi-agent evolve: Llm self-improve through co-evolution},
  author={Chen, Yixing and Wang, Yiding and Zhu, Siqi and Yu, Haofei and Feng, Tao and Zhang, Muhan and Patwary, Mostofa and You, Jiaxuan},
  journal={arXiv preprint arXiv:2510.23595},
  year={2025}
}

@article{lin2025se,
  title={Se-agent: Self-evolution trajectory optimization in multi-step reasoning with llm-based agents},
  author={Lin, Jiaye and Guo, Yifu and Han, Yuzhen and Hu, Sen and Ni, Ziyi and Wang, Licheng and Chen, Mingguang and Liu, Hongzhang and Chen, Ronghao and He, Yangfan and others},
  journal={arXiv preprint arXiv:2508.02085},
  year={2025}
}

@article{xia2025live,
  title={Live-SWE-agent: Can Software Engineering Agents Self-Evolve on the Fly?},
  author={Xia, Chunqiu Steven and Wang, Zhe and Yang, Yan and Wei, Yuxiang and Zhang, Lingming},
  journal={arXiv preprint arXiv:2511.13646},
  year={2025}
}

@article{hu2026controlled,
  title={Controlled self-evolution for algorithmic code optimization},
  author={Hu, Tu and Chen, Ronghao and Zhang, Shuo and Yin, Jianghao and Feng, Mou Xiao and Liu, Jingping and Zhang, Shaolei and Jiang, Wenqi and Fang, Yuqi and Hu, Sen and others},
  journal={arXiv preprint arXiv:2601.07348},
  year={2026}
}

@article{ding2026swe,
  title={SWE-Replay: Efficient Test-Time Scaling for Software Engineering Agents},
  author={Ding, Yifeng and Zhang, Lingming},
  journal={arXiv preprint arXiv:2601.22129},
  year={2026}
}

@article{weng2026group,
  title={Group-Evolving Agents: Open-Ended Self-Improvement via Experience Sharing},
  author={Weng, Zhaotian and Antoniades, Antonis and Nathani, Deepak and Zhang, Zhen and Pu, Xiao and Wang, Xin Eric},
  journal={arXiv preprint arXiv:2602.04837},
  year={2026}
}

@article{chen2025swe,
  title={Swe-exp: Experience-driven software issue resolution},
  author={Chen, Silin and Lin, Shaoxin and Gu, Xiaodong and Shi, Yuling and Lian, Heng and Yun, Longfei and Chen, Dong and Sun, Weiguo and Cao, Lin and Wang, Qianxiang},
  journal={arXiv preprint arXiv:2507.23361},
  year={2025}
}

@article{hao2026recreate,
  title={ReCreate: Reasoning and Creating Domain Agents Driven by Experience},
  author={Hao, Zhezheng and Wang, Hong and Luo, Jian and Zhang, Jianqing and Zhou, Yuyan and Lin, Qiang and Wang, Can and Dong, Hande and Chen, Jiawei},
  journal={arXiv preprint arXiv:2601.11100},
  year={2026}
}

@article{jimenez2023swe,
  title={Swe-bench: Can language models resolve real-world github issues?},
  author={Jimenez, Carlos E and Yang, John and Wettig, Alexander and Yao, Shunyu and Pei, Kexin and Press, Ofir and Narasimhan, Karthik},
  journal={arXiv preprint arXiv:2310.06770},
  year={2023}
}

@inproceedings{bui2022vul4j,
  title={Vul4j: A dataset of reproducible java vulnerabilities geared towards the study of program repair techniques},
  author={Bui, Quang-Cuong and Scandariato, Riccardo and Ferreyra, Nicol{\'a}s E D{\'\i}az},
  booktitle={Proceedings of the 19th International Conference on Mining Software Repositories},
  pages={464--468},
  year={2022}
}

@article{singh2025openai,
  title={Openai gpt-5 system card},
  author={Singh, Aaditya and Fry, Adam and Perelman, Adam and Tart, Adam and Ganesh, Adi and El-Kishky, Ahmed and McLaughlin, Aidan and Low, Aiden and Ostrow, AJ and Ananthram, Akhila and others},
  journal={arXiv preprint arXiv:2601.03267},
  year={2025}
}

@article{liu2024deepseek,
  title={Deepseek-v3 technical report},
  author={Liu, Aixin and Feng, Bei and Xue, Bing and Wang, Bingxuan and Wu, Bochao and Lu, Chengda and Zhao, Chenggang and Deng, Chengqi and Zhang, Chenyu and Ruan, Chong and others},
  journal={arXiv preprint arXiv:2412.19437},
  year={2024}
}

@article{yang2025qwen3,
  title={Qwen3 technical report},
  author={Yang, An and Li, Anfeng and Yang, Baosong and Zhang, Beichen and Hui, Binyuan and Zheng, Bo and Yu, Bowen and Gao, Chang and Huang, Chengen and Lv, Chenxu and others},
  journal={arXiv preprint arXiv:2505.09388},
  year={2025}
}

@article{rastogi2025devstral,
  title={Devstral: Fine-tuning Language Models for Coding Agent Applications},
  author={Rastogi, Abhinav and Yang, Adam and Jiang, Albert Q and Liu, Alexander H and Sablayrolles, Alexandre and H{\'e}liou, Am{\'e}lie and Martin, Am{\'e}lie and Agarwal, Anmol and Ehrenberg, Andy and Lo, Andy and others},
  journal={arXiv preprint arXiv:2509.25193},
  year={2025}
}

@inproceedings{yao2022react,
  title={React: Synergizing reasoning and acting in language models},
  author={Yao, Shunyu and Zhao, Jeffrey and Yu, Dian and Du, Nan and Shafran, Izhak and Narasimhan, Karthik R and Cao, Yuan},
  booktitle={The eleventh international conference on learning representations},
  year={2022}
}

@inproceedings{vrpilot,
  title={A case study of llm for automated vulnerability repair: Assessing impact of reasoning and patch validation feedback},
  author={Kulsum, Ummay and Zhu, Haotian and Xu, Bowen and d'Amorim, Marcelo},
  booktitle={Proceedings of the 1st ACM International Conference on AI-Powered Software},
  pages={103--111},
  year={2024}
}

@inproceedings{ntr,
  title={Template-Guided Program Repair in the Era of Large Language Models.},
  author={Huang, Kai and Zhang, Jian and Meng, Xiangxin and Liu, Yang},
  booktitle={ICSE},
  pages={1895--1907},
  year={2025}
}

@article{bui2024apr4vul,
  title={APR4Vul: an empirical study of automatic program repair techniques on real-world Java vulnerabilities},
  author={Bui, Quang-Cuong and Paramitha, Ranindya and Vu, Duc-Ly and Massacci, Fabio and Scandariato, Riccardo},
  journal={Empirical software engineering},
  volume={29},
  number={1},
  pages={18},
  year={2024},
  publisher={Springer}
}

@article{hu2025tsapr,
  title={TSAPR: A Tree Search Framework For Automated Program Repair},
  author={Hu, Haichuan and Shang, Ye and Sun, Weifeng and Zhang, Quanjun},
  journal={arXiv preprint arXiv:2507.01827},
  year={2025}
}

@inproceedings{10.1145/3597503.3639222,
author = {Zhou, Xin and Kim, Kisub and Xu, Bowen and Han, Donggyun and Lo, David},
title = {Out of Sight, Out of Mind: Better Automatic Vulnerability Repair by Broadening Input Ranges and Sources},
year = {2024},
publisher = {Association for Computing Machinery},
booktitle = {Proceedings of the IEEE/ACM 46th International Conference on Software Engineering},
numpages = {13},
series = {ICSE '24}
}

@article{ren2020codebleumethodautomaticevaluation,
    author = {Shuo Ren and Daya Guo and Shuai Lu and Long Zhou and Shujie Liu and Duyu Tang and Neel Sundaresan and Ming Zhou and Ambrosio Blanco and Shuai Ma},
    title = {CodeBLEU: a Method for Automatic Evaluation of Code Synthesis},
    journal = {arXiv preprint arXiv:2009.10297},
    year = 2020
}

@inproceedings{zhang2024vuladvisor,
  title={Vuladvisor: Natural language suggestion generation for software vulnerability repair},
  author={Zhang, Jian and Wang, Chong and Li, Anran and Wang, Wenhan and Li, Tianlin and Liu, Yang},
  booktitle={Proceedings of the 39th IEEE/ACM International Conference on Automated Software Engineering},
  pages={1932--1944},
  year={2024}
}

@article{roucher2025smolagents,
  title={smolagents: A smol library to build great agentic systems},
  author={Roucher, Aymeric and del Moral, A Villanova and Wolf, Thomas and von Werra, Leandro and Kaunism{\"a}ki, Erik},
  journal={Hugging Face},
  year={2025}
}

@inproceedings{xu2025amem,
  title={A-Mem: Agentic memory for llm agents},
  author={Xu, Wujiang and Liang, Zujie and Mei, Kai and Gao, Hang and Tan, Juntao and Zhang, Yongfeng},
  booktitle={Advances in Neural Information Processing Systems},
  year={2025}
}

@article{kojima2022large,
  title={Large language models are zero-shot reasoners},
  author={Kojima, Takeshi and Gu, Shixiang Shane and Reid, Machel and Matsuo, Yutaka and Iwasawa, Yusuke},
  journal={Advances in neural information processing systems},
  volume={35},
  pages={22199--22213},
  year={2022}
}

@article{brown2020language,
  title={Language models are few-shot learners},
  author={Brown, Tom and Mann, Benjamin and Ryder, Nick and Subbiah, Melanie and Kaplan, Jared D and Dhariwal, Prafulla and Neelakantan, Arvind and Shyam, Pranav and Sastry, Girish and Askell, Amanda and others},
  journal={Advances in neural information processing systems},
  volume={33},
  pages={1877--1901},
  year={2020}
}

@article{shao2026fix,
  title={Fix Pattern-Aware Vulnerability Patch Generation via In-Context Learning},
  author={Shao, Miaomiao and Ding, Yuxin and Gao, Cuiyun and Wang, Junru and Zhu, Guoqing},
  journal={ACM Transactions on Software Engineering and Methodology},
  year={2026},
  publisher={ACM New York, NY}
}

@inproceedings{ye2026well,
  title={Well Begun is Half Done: Location-Aware and Trace-Guided Iterative Automated Vulnerability Repair},
  author={Ye, Zhenlei and Sun, Xiaobing and Cao, Sicong and Bo, Lili and Li, Bin},
  booktitle = {Proceedings of the IEEE/ACM 48th International Conference on Software Engineering},
  publisher = {Association for Computing Machinery},
  year={2026},
  series = {ICSE '26}
}

@inproceedings{jin2026intent,
    title={INTENTFIX: Automated Logic Vulnerability Repair via LLM-Driven Intent Modeling},
    author={Jinseok, Heo and Dongwook, Choi and Jinyoung, Kim and Misoo, Kim and  Eunseok, Lee},
    booktitle = {Proceedings of the IEEE/ACM 48th International Conference on Software Engineering},
    publisher = {Association for Computing Machinery},
    year={2026},
    series = {ICSE '26}
}

@article{liu2026contrafix,
  title={ContraFix: Agentic Vulnerability Repair via Differential Runtime Evidence and Skill Reuse},
  author={Liu, Simiao and Liu, Fang and Zhang, Li and Liu, Yang and Zhu, Yinghao},
  journal={arXiv preprint arXiv:2605.17450},
  year={2026}
}

@misc{anthropic2025claudecode,
  title={Claude Code},
  author={Anthropic},
  year={2025},
  howpublished={\url{https://github.com/anthropics/claude-code}}
}

@inproceedings{cheng2025automated,
  title={Automated vulnerability repair based on retrieval-augmented generation},
  author={Cheng, Shengyi and Yu, Qiao and Zhu, Yi and Huang, Zirui},
  booktitle={2025 7th International Conference on Information Science, Electrical and Automation Engineering (ISEAE)},
  pages={941--947},
  year={2025},
  organization={IEEE}
}

@article{shinn2023reflexion,
  title={Reflexion: Language agents with verbal reinforcement learning},
  author={Shinn, Noah and Cassano, Federico and Gopinath, Ashwin and Narasimhan, Karthik and Yao, Shunyu},
  journal={Advances in neural information processing systems},
  volume={36},
  pages={8634--8652},
  year={2023}
}

@article{majumder2023clin,
  title={Clin: A continually learning language agent for rapid task adaptation and generalization},
  author={Majumder, Bodhisattwa Prasad and Mishra, Bhavana Dalvi and Jansen, Peter and Tafjord, Oyvind and Tandon, Niket and Zhang, Li and Callison-Burch, Chris and Clark, Peter},
  journal={arXiv preprint arXiv:2310.10134},
  year={2023}
}

@inproceedings{hu2026psearch,
  title={PSearch: Search-Based Patch Generation in the Era of LLM-Based Automated Program Repair},
  author={Hu, Haichuan and Shang, Ye and Sun, Weifeng and Zhang, Quanjun},
  booktitle={Proceedings of the 41st IEEE/ACM International Conference on Automated Software Engineering},
  pages={1157--1169},
  year={2026}
}

@inproceedings{hu2026multifixer,
  title={MultiFixer: A Coordinator-Proposer Based Multi-Agent Framework For Fixing Multi-Hunk Bugs},
  author={Hu, Haichuan and Fang, Chunrong and Shang, Ye and Liu, Jiawei and Sun, Weifeng and Xie, Guoqing and Zhong, Chenxing and Zhang, Quanjun},
  booktitle={Proceedings of the 41st IEEE/ACM International Conference on Automated Software Engineering},
  pages={2934--2946},
  year={2026}
}

@article{zhou2026self,
  title={Self-Evolving Coding Agents},
  author={Zhou, Hao and Hu, Haichuan and Luo, Tianyu and Shang, Ye and Fang, Chunrong and Chen, Zhenyu and Xiao, Liang and Zhang, Quanjun},
  journal={arXiv preprint arXiv:2608.03392},
  year={2026}
}

@inproceedings{hu2025can,
  title={Can gpt-o1 kill all bugs? an evaluation of gpt-family llms on quixbugs},
  author={Hu, Haichuan and Shang, Ye and Xu, Guolin and He, Congqing and Zhang, Quanjun},
  booktitle={2025 IEEE/ACM International Workshop on Automated Program Repair (APR)},
  pages={11--18},
  year={2025},
  organization={IEEE}
}

@article{hu2025repair,
  title={Repair-r1: Better test before repair},
  author={Hu, Haichuan and Xie, Xiaochen and Zhang, Quanjun},
  journal={arXiv preprint arXiv:2507.22853},
  year={2025}
}

@article{hu2026beyond,
  title={Beyond the Model: Demystifying Harness Effects in Software Engineering Agents},
  author={Hu, Haichuan and Zhang, Quanjun and Yu, Shengcheng and Chen, Zhifei and Luo, Tianyu and Fang, Chunrong and Chen, Zhenyu and Xiao, Liang},
  journal={arXiv preprint arXiv:2609.32459},
  year={2026}
}

@article{zhang2026effectiveness,
  title={On the Effectiveness of Code Representation in Deep Learning-Based Automated Patch Correctness Assessment},
  author={Zhang, Quanjun and Hu, Haichuan and Fang, Chunrong and Shang, Ye and Zheng, Tao and Chen, Zhenyu and Yang, Yun and Xiao, Liang},
  journal={arXiv preprint arXiv:2603.07520},
  year={2026},
  doi={10.48550/arXiv.2603.07520}
}

@article{zhang2024machine,
  title={Machine translation testing via syntactic tree pruning},
  author={Zhang, Quanjun and Zhai, Juan and Fang, Chunrong and Liu, Jiawei and Sun, Weisong and Hu, Haichuan and Wang, Qingyu},
  journal={ACM Transactions on Software Engineering and Methodology},
  volume={33},
  number={5},
  pages={1--39},
  year={2024},
  publisher={ACM, New York, NY}
}

@article{zhang2026compass,
  title={ComPass: Contrastive Learning for Automated Patch Correctness Assessment in Program Repair},
  author={Zhang, Quanjun and Shang, Ye and Hu, Haichuan and Fang, Chunrong and Chen, Zhenyu and Xiao, Liang},
  journal={Empirical Software Engineering},
  volume={31},
  number={6},
  pages={157},
  year={2026},
  publisher={Springer}
}
